\documentclass[epj,nopacs]{svjour}
\usepackage{amsmath,amssymb,amsfonts,graphicx,cite,multirow,color,algorithm}
\usepackage{algpseudocode}
\usepackage[utf8]{inputenc}
\usepackage{amsmath}
\usepackage{caption}
\usepackage{subcaption}
\usepackage{hyperref}
\usepackage{tikz}
\usepackage{grffile}
\usetikzlibrary{calc,math,decorations.markings}
\usepackage{soul}
\graphicspath{{graphics/}}
\makeatletter
\ifx\input@path\@undefined
\def\input@path{{graphics/}}
\else
\g@addto@macro\input@path{{graphics/}}
\fi
\makeatother

\newcommand{\code}[1]{\texttt{#1}}

\newcommand{\meas}{\mathrm{d}}
\newcommand{\measDiff}{\mathrm{d}}
\newcommand{\nix}[1]{}
\usepackage{snapshot}
\usepackage{ulem}

\DeclareMathSizes{10}{10}{7}{4} 
\DeclareMathSizes{10}{10}{7}{3}

\preprint{MCnet-25-10, KA-TP-14-2025}

\title{Phenomenological constraints of the building blocks of the
  cluster hadronization model}

\author{Stefan Gieseke\inst{1}, Stefan Kiebacher\inst{1}, Simon
  Pl\"atzer\inst{2,3}, Jan Priedigkeit\inst{2}}

\institute{Institute for Theoretical Physics (ITP), KIT,
  Wolfgang-Gaede-Straße 1, D-76128 Karlsruhe \and Institute of Physics,
  NAWI Graz, University of Graz, Universit\"atsplatz 5, A-8010 Graz,
  Austria \and Particle Physics, Faculty of Physics, University of
  Vienna, Boltzmanngasse 5, A-1090 Wien, Austria}

\authorrunning{S.~Gieseke, S.~Kiebacher, S.~Pl\"atzer, J.~Priedigkeit}
\titlerunning{Dissecting the cluster model}

\date{\today}

\abstract{We introduce building blocks for the cluster hadronization model in
  light of a new structure, focusing on cluster fission and cluster decay. We
  propose theoretically motivated matrix elements for cluster fission and decay
  as building blocks and study some first phenomenological implications at
  different energies. In particular we develop a set of observables which can
  be used to dissect the hadronization history and have constraining power on
  the individual building blocks. Our analysis will be completed by including
colour reconnection in a follow-up work.}

\begin{document}

\maketitle


\section{Introduction}
\label{sec:Introduction}

The Large Hadron Collider (LHC) has exceeded expectations about the
precision at which hadronic observables can be measured at a hadron
collider.  One key role for the understanding of fully exclusive
hadronic final states is attributed to the simulation with Monte Carlo
event generators such as \code{Herwig}
\cite{Bellm:2015jjp,Bellm:2017bvx,Bellm:2019zci,Bewick:2023tfi},
\code{Pythia} \cite{Bierlich:2022pfr}, or \code{Sherpa}
\cite{Sherpa:2024mfk}. Phenomenological models enter these simulations
to describe the hadronization of the partonic final state and the
Multiple Partonic Interactions (MPI). For many observables the
uncertainties associated with their modelling are among the largest
parts of the systematic errors that are attributed to precision
observables. While in the last decades a strong focus in the
development of event generators has been the matching and merging of
higher order perturbative matrix elements with the parton shower. The
parton showers themselves have gained attention only in recent years,
with developments demonstrating shower accuracy at higher logarithmic accuracy
\cite{Dasgupta:2020fwr,Forshaw:2020wrq,Nagy:2020rmk,Herren:2022jej,vanBeekveld:2022zhl,Hoche:2024dee,vanBeekveld:2024wws},
and beyond the limit of leading-colour evolution
\cite{Platzer:2013fha,Forshaw:2019ver,DeAngelis:2020rvq,Forshaw:2025bmo}.
Hadronization models, however, have received less attention. Only recently it
has become clear that parton showers and \linebreak hadronization can not be viewed in
isolation and that they necessarily need to interrelate in light of the
requirements of infrared safety and factorized calculations
\cite{Platzer:2022jny}.

In this paper we want to address some of the structure of the cluster
hadronization model in \code{Herwig} and point out possible
perturbative and non-perturbative input to construct a novel model.
We in particular focus on observables which allow to pinpoint a
number of deficiencies in the description of the hadronic final state,
and the possibility to constrain the building blocks of the model in great
detail. While inclusive observables and event shapes from the LEP era
have so far been the most important input to the tuning of
hadronization models, we here go beyond and demonstrate that novel
observables can allow us to take a deeper look into the structure of
the hadronization model, thus complementing the theoretical development
behind factorized approaches and evolution equations \cite{Platzer:2022jny}.  There
are new data from the Large Hadron Collider (LHC) and from legacy
$e^+e^-$ data for detailed phase space distributions of identified
particles and their correlations, which show that some of the details
in the cluster hadronization model are not adequate for their
description.  Here we need improvements of the detailed modelling of
hadron production from the cluster model.

Our main approach is based on the observation that parton showers and
hadronization models cannot be \linebreak looked at in isolation
\cite{Platzer:2022jny,Hoang:2024zwl}, and that the hadronziation model
needs to follow a smooth onset from the partonic state generated by
the parton shower.  Within this approach, the parton shower cutoff is
viewed as a factorization scale. It will therefore not strictly serve
as a scale below which partons become unresolved or the meaning of
partons as quanta of a weakly coupled field becomes meaningless
overall. We should rather argue that this scale should at the same
time be viewed as an upper bound for the evolution of a hadronization
model, using effectively low-scale perturbative processes together
with a non-perturbative initial condition, rather than a mere endpoint
of the partonic picture.  With this idea in mind we will show that
further details of the dynamics in the hadronization model can be
derived or at least be motivated. Another shortcoming is the close
interplay of hadronization and colour-reconnection, which is thus far
modelled on a phenomenological basis and not viewed as an integral
part of a hadronization model, but for which an equally fundamental
motivation exists \cite{SGE_CR,Platzer:2022jny}.

In the present work, we focus on the cluster fission and decay
dynamics and identify possible perturbative and non-perturbative input
to their role in an evolving hadronization model. We will in
particular address the roles of phase space and kinematics, as well as
the elementary (perturbative) processes which drive the cluster fission
process and the final cluster decay to hadrons. We will then also
focus on identifying a set of observables which we can use to
complement such a design in order to constrain the building blocks and
their interplay in the according measurements. A similar analysis for
colour reconnection will be presented in a follow-up work, together
with a first tuning analysis. In particular, our new model will lead
to {\it less parameters} to be considered in such a tuning, while it
introduces more dynamical features. Our work will constitute a key
cornerstone to develop and scrutinize a fully evolving hadronization
model and to shed light on its infrared cutoff dependence.

The outline of this paper is as follows: In
Sec.~\ref{sec:ClusterModel}, we present the main framework in which we
will analyze and improve the cluster hadronization model. This
includes a review of its current state, and the construction of its
building blocks in particular towards the kinematic and phase space
constraints of the cluster fission and decay processes. In
Sec.~\ref{sec:Benchmarks} we study the impact of the modifications
onto measured observables, in particular extending the energy range
which should be considered to pinpoint low-scale dynamics of the
interface between parton showers and hadronization. In
Sec.~\ref{sec:NewObservables} we develop an entire set of observables
which can be used to disentangle the various building blocks we
propose, and to guide further theoretical development. We then
conclude and provide an outlook on the role of colour reconnection, as
well as further development of a novel model.

\section{Clusters in light of perturbative evolution}
\label{sec:ClusterModel}

Demanding that the hadronization model should smoothly match to the
perturbative evolution of the parton shower has been proposed in
\cite{Platzer:2022jny}, and in \cite{Hoang:2024zwl}, where the cluster
fission process, as well as the low-scale splitting of gluons have
been viewed as a process which should ultimately been driven by the
parton shower dynamics such that a smooth transition can be
guaranteed. We will further elaborate on such an approach, and discuss
in this section how the cluster model can be generalized and what
perturbative building blocks we can identify in its construction, in
particular for the cluster fission. We also propose, what
non-perturbative dynamics we could assume to govern the cluster
decays. Before doing so, we review the structure of the cluster model
as it is currently available. This structure will ultimately need to
be changed, in particular to respect the infrared safety criterion
spelled out in \cite{Platzer:2022jny} and as such the demand that
clusters should include gluons as constituents, however at this point
we do not rebuild the entire structure of the cluster model.

\subsection{General structure of the cluster model}

\begin{figure}[t]
  \includegraphics[width=0.49\textwidth]{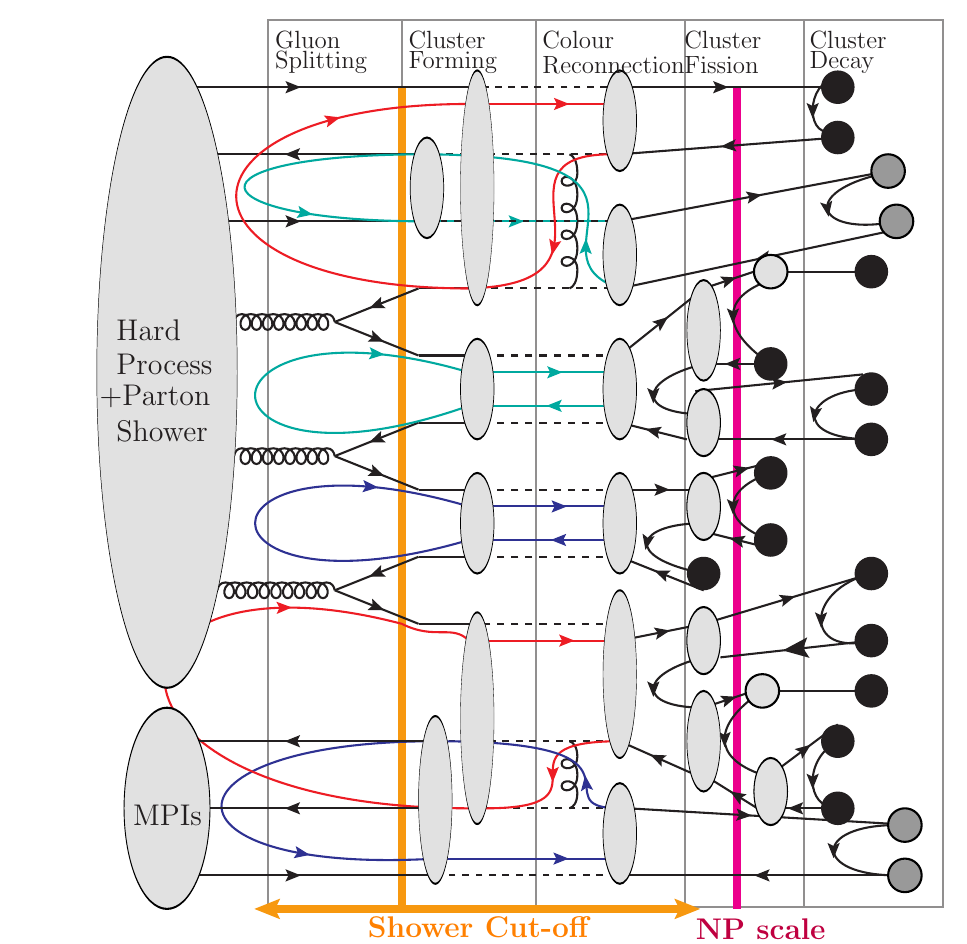}
  \caption{The structure of the cluster hadronization model. We
    illustrate how clusters (grey blobs) are formed after the parton
    shower and the low-scale gluon splitting, how they fission and how
    they reconnect ({\it i.e.} how they rescatter), and how they
    ultimately decay into stable or unstable hadrons (black and dark
    grey circles). We also indicate some characteristic scales, in
    particular the parton shower cutoff, and a final, non-perturbative
    cutoff at which the initial condition of the model is fixed to
    consist of conversions of clusters into hadrons.}
  \label{fig:HadronizationPicture}
\end{figure}

In Fig.~\ref{fig:HadronizationPicture} we show a picture of the
hadronization as it evolves from the parton shower.  The hadronization
generally starts from a partonic final state where all partons are in
a certain colour state.  Usually the colour state is defined in the
limit of large number of colours, $N_c$, such that all quarks and
antiquarks are uniquely assigned to the endpoint of a colour line.
Gluons carry both a colour and an anticolour.  In order to assign all
partons to a colour singlet state in a unique way, all particles, in
particular the gluons, are assigned a constituent mass and undergo a
number of (small) boosts in order to maintain momentum conservation.
This will allow us to decay the gluons non-perturbatively into a light
$q\bar q$-pair.  The gluon mass is just above the threshold for the
creation of this $q\bar q$-pair which implies that even with an
isotropic decay the resulting quarks move practically in the original
direction of the gluon and share its momentum almost equally and carry
the respective colour line of the gluon.  As a result, the final state
consists of quarks, antiquarks and possibly (anti-)diquarks from the
initial hadrons.  All particles are in a colour triplet or antitriplet
state and rest at the end of a colour line.  The colour partner of
each particle at the other end of the colour line can be easily found
and both form one of the so-called primordial clusters.  This set of
primordial clusters will be the starting point of our studies.

A cluster $C(q_1,q_2)$ is composed of two constituents $q_1$ and $q_2$
which generally are one quark or anti-diquark and one anti-quark or
diquark and carries the net momentum of these two constituents,
resulting in an invariant mass squared $M^2=(p_1+p_2)^2$, while the
constituents are on their constituent mass-shell.  The invariant mass
$M$ of the cluster now determines the further steps towards
hadronization.  Clusters with a large mass $M$, i.e.\ $M$ larger than
typically $3-4$ GeV, will fission into pairs of lighter clusters until
all clusters are light.  We call this process cluster fission.
Once all cluster masses are below a certain threshold they will decay
into pairs of hadrons, we call this cluster decay.  It is
possible that a cluster ends up with a mass $M$ that does not leave
enough phase space to decay even into the lightest possible pair of
hadrons.  In this exceptional case, the cluster will decay into a single
hadron, while momentum is exchanged with a neighbouring cluster in order
to maintain momentum conservation.  

\begin{figure}[t]
  \centering
  \includegraphics[width=0.4\textwidth]{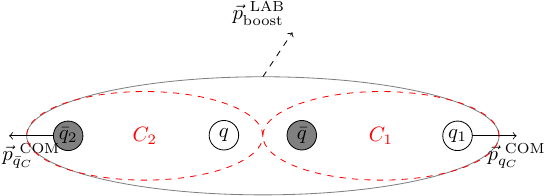}
  \includegraphics[width=0.4\textwidth]{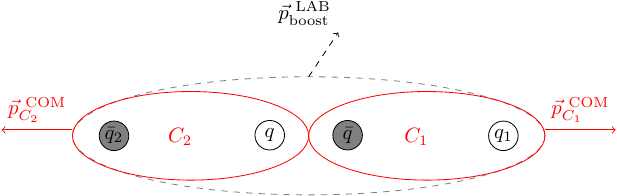}
  \caption{Sketch of current cluster fission model in \code{Herwig}.}
  \label{fig:alignedFission}
\end{figure}

The threshold below which clusters decay to hadrons is usually a fixed
cutoff, though it has recently been proposed to accompany this
parameter by a further smearing. In light of our aim to reduce the
number of parameters, while implementing dynamics which follow from a
clean theoretical analysis we refrain to use this smearing and
interpret the fixed value as the unique point in the evolution when
the initial condition has been fixed. In Fig.~\ref{fig:alignedFission}
we sketch a typical cluster fission kinematics.
A cluster fission is always carried out when the cluster mass fulfils
\begin{align}
  M \geq \left[{M_{\rm max}}^{a}+(m_1+m_2)^{a}\right]^{1/a} \ ,
\end{align}
where $M_{\rm max} = \textbf{Clmax}$ and $a = \textbf{Clpow}$
are tunable parameters of the
hadronization model and $m_{1, 2}$ are the constituent masses of the
quarks $q_{1,2}$ inside the cluster in question.  In this model, the
kinematics of cluster fission
\begin{align}
  \label{eq:fission}
  C({q}_1,{q}_2) \rightarrow C_1({q}_1,{q}),C_2(\bar{q},{q}_2)
\end{align}
is determined by restricting the fission products to be aligned with the
cluster axis along which the original \linebreak quarks move back-to-back in the
cluster rest frame.  In this fission a light $q\bar q$ pair with flavour
$q\in \{u, d, s\}$ is drawn according to constant tunable relative
weights $\textbf{Pwt}_q$ and the masses $M_1$ and $M_2$ of the children
$C_{1,2}$ are usually drawn according to a power law,
\begin{align}
  \label{eq:m12power}
  (M_1 M_2)^{p-1}\ ,
\end{align}
where $p=\textbf{PSplit}$ is a further parameter of the model.  The masses
$M_{1,2}$ must of course be kinematically allowed, as the constituents
carry masses.  The value of $p$ determines whether the children are
rather heavy and slow, possibly leading to a longer chain of cluster
fission until the threshold is reached, or whether a cluster quickly
fissions into light and fast clusters.  This will have a strong impact
on the resulting multiplicity and spectra of hadrons.  Once the masses
$M_{1,2}$ are known, the momenta of all constituents can be determined,
as also the new pair of quarks $q$ is assumed to move along the cluster
axis.  Note that the phenomenological parameters $M_{\rm max}$, $a$ and
$p$ are each a set of three parameters in \code{Herwig}, depending on
whether the original cluster has a only $\{u,d,s\}$, $c$ or $b$ as
constituents.

Once all clusters are fissioned they decay into pairs of hadrons.  Here,
once more a $q\bar q$ pair is drawn from the vacuum, such that two hadrons
can form in pairs $H_1(q_1, q)$ and $H_2(\bar q, q_2)$.  Note that also
diquarks occur and hence the hadrons could be either mesons or baryons.
All kinematically possible pairs of available hadrons are determined for
every cluster and are assigned a weight that includes weight factors for
angular momentum, flavour multiplet and phase space.  Finally a pair is
drawn according to these weights and the cluster decays isotropically
into this hadron pair in its rest frame. Further details are
described here \cite{Bahr:2008pv}. 

In this work, we want to address two particular issues of this model.
The first part is that the kinematics of cluster fission is strictly
longitudinal along the cluster axis.  On the other hand, the cluster
decay is isotropic.  This results in a hard transition from
longitudinal to isotropic kinematics.  Small transverse momentum
smearing is only introduced via the cluster decay.  If we view the
cluster fission as a low-scale perturbative continuation of the parton
shower this is once again a rather hard transition from soft and
collinear dynamics to strictly longitudinal dynamics.  Originally,
these assumptions were made for simplicity in order not to attribute
effects that should be determined by perturbative parton dynamics to
the hadronization model.  We will discuss these issues in the
remainder of this work.

\subsection{Hints from data}

With new data on the production of identified particles we find hints
that the hadronization model in its current implementation might have
shortcomings.  One example is the two-particle angular correlation of
baryons as measured by the ALICE collaboration
\cite{ALICE_pp_correlations_ALICE_2016_I1507157,ALICE_Xi_correlations}.

Baryon production in the cluster hadronization model is usually
achieved with the production of diquarks in intermediate steps.
Originally, the only source of baryons, besides the remnants of the
incoming hadrons, which are a somewhat special case and not addressed
here in particular, was the cluster decay into hadron pairs.  Here, a
mesonic cluster, i.e.\ a cluster made of a quark-antiquark pair, could
decay into a pair of baryons by producing an intermediate pair of a
diquark and anti-diquark of opposite flavour.  As a result, baryon
pairs are always strongly correlated in phase space as they always
originate from the same cluster.  A newer treatment of
colour-reconnection in \code{Herwig} \cite{Gieseke:2017clv} has lead
to some modification of this effect.  Here, baryonic clusters,
i.e.\ clusters made up of three quarks or three antiquarks, could
arise from the colour rearrangement of nearby mesonic clusters.  This
leads to some decorrelation of baryon pairs.

\begin{figure}
  \centering
  \includegraphics[width=0.5\textwidth]{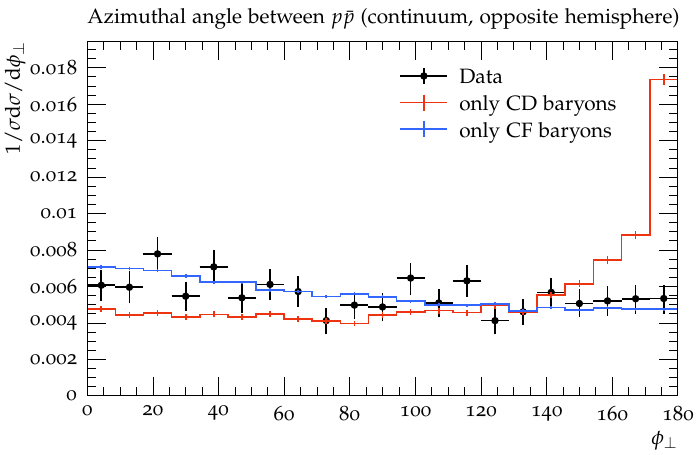}
  \includegraphics[width=0.5\textwidth]{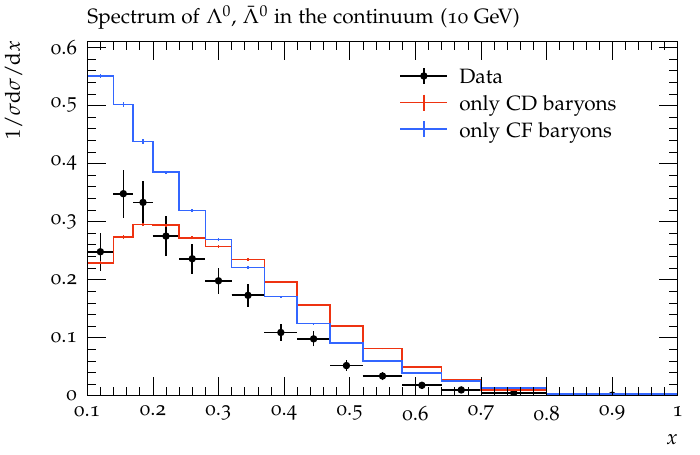}
  \caption{The $p\bar{p}$ azimuthal angular correlations for $e^+e^-$
    collisions at $\sqrt{s}=10$ GeV measured by ARGUS
    \protect\cite{ARGUS_ppbar_corr} for the opposite hemisphere (upper
    panel). $\phi_T$ is the angle between the transverse momenta with
    respect to the thrust axis,
    $\cos \phi_T =
    \hat{p}_{\perp,p}\cdot\hat{p}_{\perp,\bar{p}}$. $\Lambda^0 $ baryon
    distribution of scaled momentum, $x=2|\vec{p}|/\sqrt{s}$, measured
    by ARGUS \protect\cite{ARGUS_hyperon_spectra} (lower panel). }
  \label{fig:ARGUS_correlations_and_spectrum_CFCDbaryons}
\end{figure}

However, this is not the final answer, as the correlation observables
are found to also depend strongly on details of the cluster fission
and decay.  As an example we show in
Fig.~\ref{fig:ARGUS_correlations_and_spectrum_CFCDbaryons} ARGUS data
at $\sqrt{s}=10$ GeV $e^+e^-$ collisions for baryon production only
during the cluster decay (the default in \code{Herwig}) and only
during the cluster fission. While the $p\bar{p}$ correlation data
(upper panel) strongly favours cluster fission baryons, the baryon
momentum spectrum data (lower panel) strongly favours the cluster
decay baryons. These findings generalize also to high energy hadronic
collisions. Therefore a deeper investigation of the interplay between
cluster fission, cluster decay and colour reconnection is needed.

We will address the aforementioned colour reconnection effect,
together with other effects in a forthcoming paper, which will
complement the current study.  This intertwined view on colour
reconnection and dynamics of the cluster model already shows that it
might be advantageous to consider these two effects in a more
universal model.  Such a model would address the non-perturbative
evolution of momentum and colour degrees of freedom simultaneously.
In this light, the current work should be regarded as another step
towards a more universal model, and to provide hints to identify a
number of observables which could constrain such an approach.

In this work we will address two closely related issues which both are
the dynamical formulation of cluster fission and decay.  In order to
be able to address these processes in an unbiased way, we begin by
disentangling the phase space for these processes from the actual
dynamics that should be formulated as Lorentz-invariant matrix
elements in both cases. For the cluster fission we will assume a
perturbatively inspired matrix element that continues the perturbative
evolution of the parton shower. The cluster decay ultimately needs to
be derived within a framework of hadronic matrix elements. In our
present work, we consider a more elaborate model than the isotropic
decay in that we assume that the cluster to hadron transition is
effectively governed by a $t$-channel type exchange with a
non-perturbative matrix element.

As a guideline for the formulation of the model we keep in mind that
the transitions between the different parts of the model should be
smooth in order to reduce the dependence on intermediate scales.  The
cluster fission should be viewed as an evolution from an upper scale,
the parton shower cutoff scale $\mu^2$ towards a lower hadronic scale,
at which we transition to an initial condition which ultimately can
only be determined non-perturbatively. We emphasise that this implies
that if the parton shower evolution is stopped early, the cluster
fission must act as a smooth continuation of the soft and collinear
dynamics in the parton shower \cite{Platzer:2022jny,Hoang:2024zwl}, as
illustrated in Fig.~\ref{fig:HadronizationPicture}.  Likewise, the
cluster decay should pick up at the lower end of the cluster fission
but has a discrete spectrum of hadron masses as lower bound and is
therefore naturally regularized in its infrared region.

\subsection{Phase space and matrix elements for cluster fission}

\begin{figure}[t]
  \centering
  \includegraphics[width=0.45\textwidth]{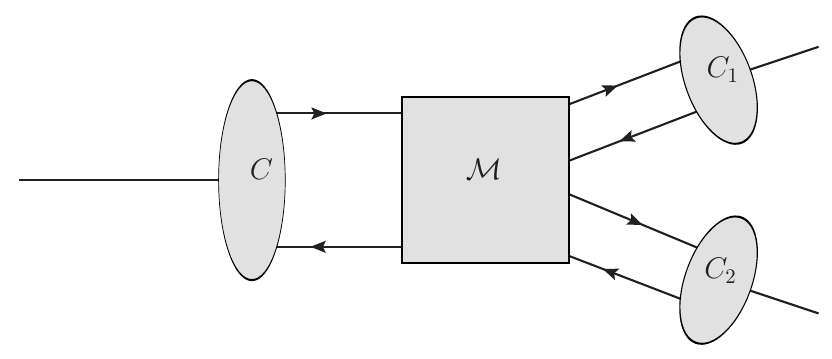}
  \caption{Diagram of Cluster Fission}
  \label{fig:BlobCF}
\end{figure}

In order to generically describe cluster fission in the aforementioned
matrix element picture, we first employ the factorization of phase
space in order to separate the pure phase space effects from the
actual matrix element as depicted in Fig.~\ref{fig:BlobCF}. Those
result from two assumptions by simply dissecting the relevant Feynman
graphs: Clusters, as imprinted by the nature of the external hadron
amplitudes, essentially provide us with a 2-Particle-Irreducible (2PI)
picture \cite{Platzer:2022jny} in which we can organise the underlying
Feynman graph using two-particle interaction diagrams as propagators
of composite systems, which would also be removed in truncating for
the use of external wave functions. We can then make the assumption
that we effectively expand around on-shell quarks and gluons (on their
constituent mass shell), while we currently neglect the four-point
functions which evolve clusters into themselves.

We can write down a general $2\rightarrow 4 $ partonic process with a
matrix element
$\mathcal{M}_{2\rightarrow 4}(p_1,p_2\rightarrow q_1,q,\bar{q},q_2)$,
where the rate $\Gamma_{2\rightarrow 4}$ can be written in this manner
in $D=d+1$ space-time dimensions:
\begin{align}
  \label{eq:genPS}
  \Gamma_{2\rightarrow 4} &=\int |\mathcal{M}_{2\rightarrow 4}|^2 \meas \Phi_4 \\
  \meas \Phi_4 &=\meas \Pi_{q_1}\meas \Pi_{q}\meas \Pi_{\bar{q}}\meas \Pi_{q_2}(2\pi)^D\delta^{D} (p_1+p_2-\sum_f q_f)\\
  \meas \Pi_{p_i} &=\frac{\meas^{d} p_i}{(2\pi)^{d}~2E_{p_i}}
\end{align}
Note that in the above expression we choose the masses of the quarks to
be the constituent masses instead of the current masses. This assumption
was made since the cluster fission happens at a stage, where the current
mass description breaks down. Otherwise, the cluster fission could lead
to extremely light clusters.
\begin{figure}[t]
  \centering
  \includegraphics[width=0.4\textwidth]{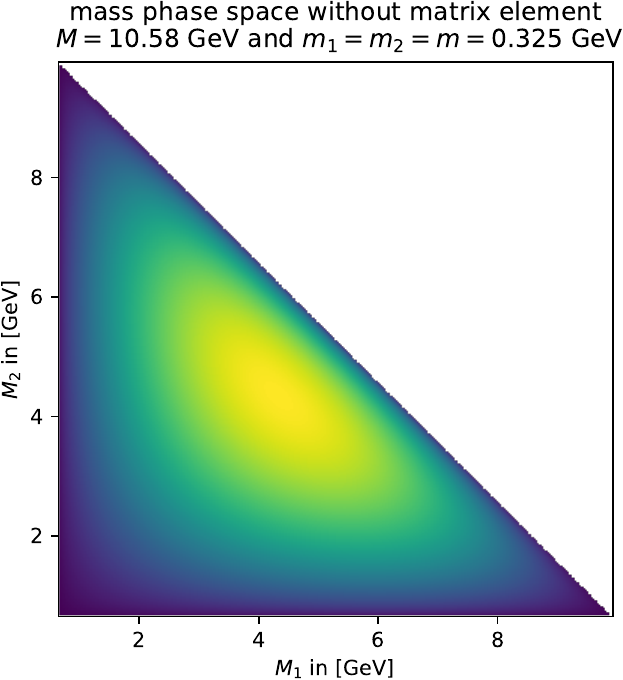}%
  \caption{Plot of flat mass phase space according to Eq.~\eqref{eq:MassPhaseSpace}.}
  \label{fig:FlatMassSpace}
\end{figure}
Rearranging the above expression
we arrive at the following equation:
\begin{align}
  \label{eq:Gamma24}
  &\Gamma_{2\rightarrow 4} \equiv
  \int \meas M_1 \meas M_2 f_\text{PS}(M_1,M_2) \nonumber\\ 
  &\qquad\qquad \meas\Omega^\text{COM}\meas\Omega^\text{COM,1}
    \meas\Omega^\text{COM,2}|\mathcal{M}_{2\rightarrow 4}|^2,\\
  &f_\text{PS} (M_1,M_2) = \nonumber\\
  &\quad f_0 \frac{\sqrt{\lambda(M^2,M_1^2,M_2^2)\lambda(M_1^2,m_1^2,m_q^2)%
    \lambda(M_2^2,m_2^2,m_q^2)}}{2M_1 2M_2}\ ,
    \label{eq:MassPhaseSpace}
\end{align}
with a constant normalization factor $f_0$ and the Källén function $\lambda(x,y,z)$.
Here, $f_\text{PS}
(M_1,M_2)$ describes the mass distribution of clusters if the matrix
element would have no implicit or explicit dependence on $M_1,M_2$,
i.e.\ the distribution coming purely from the $4$-particle phase
space. We define $\Omega^\text{COM}$ as the solid angle between the
original constituent direction $p_1$ and the child cluster
$Q_1=(q_1+q)$ in the rest frame of the original cluster. The solid
angles $\Omega^\text{COM,i}$ are the angles between the child cluster
$Q_i$ direction in the rest frame of the original cluster and $q_i$ in
the rest frame of $Q_i$ respectively.  We display the distribution
$f_\text{PS} (M_1,M_2)$ in Fig.~\ref{fig:FlatMassSpace} for
$m_1=m_2=m\equiv m_{u/d}$ at $M=\sqrt{s}_\text{BELLE}$, where $
m_{u/d}$ is the constituent mass value of ${u/d}$ quarks and
$\sqrt{s}_\text{BELLE}$ is the COM energy of the BELLE experiment as
we will later compare to BELLE data \cite{BELLEexpGeneral}.  One can
clearly note that the distribution in Fig.~\ref{fig:FlatMassSpace} is
very different from a simple power law due to the threshold effects of
the Källén functions in Eq.~\eqref{eq:MassPhaseSpace}.

Note that Eq.~\eqref{eq:Gamma24} with a matrix element
$|\mathcal{M}_{2\rightarrow 4}|^2=1$ would sample the directions of
clusters and their constituents completely isotropically.  We rather
want to extend the original picture of a strictly longitudinal cluster
fission towards a picture that smoothly emerges from collinear
emissions in the parton shower.

\subsection{The Matrix Element}
\begin{figure}[t]
  \centering
  \includegraphics[width=0.5\textwidth]{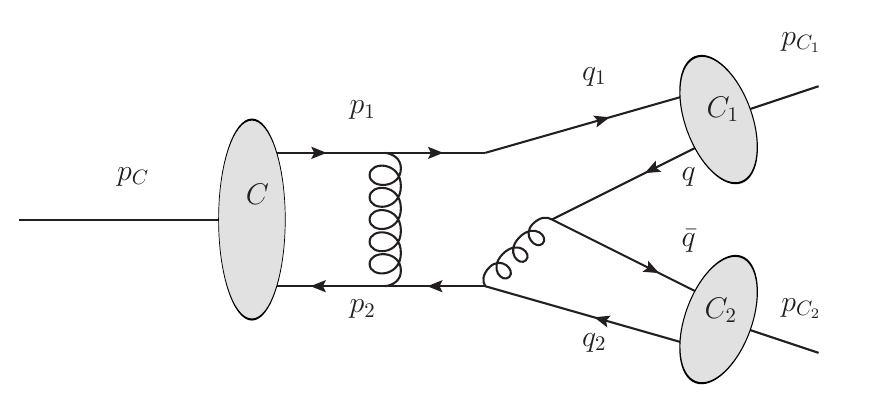}
  \includegraphics[width=0.5\textwidth]{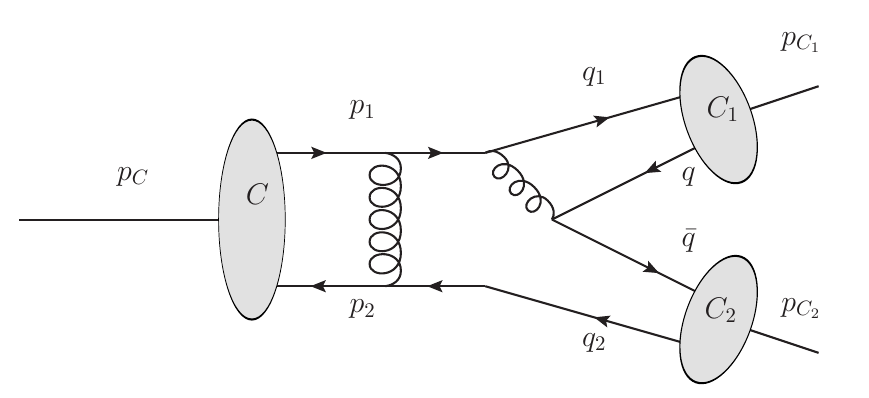}
  \caption{Feynman diagrams of the chosen cluster fission matrix element.}
  \label{fig:FeynmanDiagram}
\end{figure}
In order to achieve a non-trivial description of cluster fission that
reflects the original picture of longitudinal fission as well as a
smooth continuation of the parton shower evolution picture we make the
following ansatz.  As a cluster fission matrix element we choose a
tree-level soft $q\bar{q}$-emission, where a $t$-channel gluon
exchange is needed in order to conserve momentum as shown in
Fig.~\ref{fig:FeynmanDiagram}. A different way of viewing this
exchange is the 2PI picture: we attach a cluster propagator in the
lowest order together with the cluster branching matrix
element. In order to simplify the matrix element, we employ the infra-red
factorization from \cite{Catani} to factorize the $t$-channel from the soft
function $S$, yielding in total the following:
\begin{align}
  |\mathcal{M}_{2\rightarrow 4}|^2 &= A_0 |\mathcal{M}_{2\rightarrow
    2}|_{t}^2 S(q_1,q_2,q,\bar{q})\\ |\mathcal{M}_{2\rightarrow
    2}|_{t}^2 &= \frac{(q_1\cdot q_2)(p_1\cdot p_2)+(q_1\cdot
    p_2)(p_1\cdot q_2)}{\left[(p_1 - q_1)^2-\epsilon
      m_g^2\right]\left[(p_2 - q_2)^2-\epsilon m_g^2\right]}
\end{align}
It should be stressed that momentum conservation cannot usually be
regarded a subleading effect within the realm of the hadronization
model, and formally we should in fact consider next-to-leading power
expansions of the factorized formulae. In the present work we do not
consider the matrix element at this detailed level but rather make the
assumption that energy-momentum conservation is properly implemented,
and that the matrix element can actually be expressed in terms of
particles ``before'' and ``after'' the emission, despite that the soft
limit would normally not provide us with such an expression. A full
derivation, including subleading-power effects, would be possible
within the framework which has been sketched in
\cite{Platzer:2022nfu}, which is explicitly accounting for recoil and
mass schemes within the factorization of matrix elements.

In addition to this, we need to address the infra-red divergence of
the $t$-channel, which should be cancelled by the corresponding
emission of a soft gluon of a $p_1p_2$-dipole. For the cluster model
we put the gluons after the parton shower on a gluon constituent mass
shell $m_g$.  Hence, a reasonable regulator for our diagram shall be
this gluon mass alongside with a parameter $\epsilon$, which we choose
to be $1$ for all further considerations. In principle one could check
the validity of the cancelling IR divergences, by varying the gluon
constituent mass by a factor, but in practise this could inhibit some
$g\rightarrow s\bar{s}$ splittings.  Regarding future work, we want to
mention that we will capture a partial effect of soft-gluon evolution
colour reconnection, which needs to be subtracted properly from the
colour reconnection if one desires to interleave the two
\cite{SGE_CR}. The soft function $S(q_1,q_2,q,\bar{q})$ consists of
the following eikonal functions $I_{ij}$, which are defined as the
following in \cite{Catani}:
\begin{align}
  &I_{ij}(p_i,p_j,q,\bar{q}) =\nonumber\\ &\quad \frac{(p_i\cdot
    q)(p_j \cdot \bar{q})+(p_j\cdot q)(p_i\cdot \bar{q})-(p_i\cdot
    p_j)(q\cdot \bar{q})}{(q\cdot\bar{q})^2\left[p_i\cdot (q
      +\bar{q})\right]\left[p_j\cdot (q
      +\bar{q})\right]}\\ &S(q_1,q_2,q,\bar{q}) = (I_{11}+I_{22}-2
  I_{12})
\end{align}
Note that we again explicitly break momentum conservation since in the
soft limit $p_i\rightarrow q_i$, but in practise we have $p_i\neq
q_i$. For the case at hand we choose $q_1,q_2$ for the soft function
since we want to consider the matrix element shown in
Fig.~\ref{fig:FeynmanDiagram}, where the $q\bar{q}$-pair is emitted from
the final state. The other convention was tried, but did not
significantly impact the results. Furthermore we restored the mass
dependence of the soft function by replacing
$q\cdot\bar{q}\rightarrow q\cdot\bar{q}+m^2$, where $m^2=q^2=\bar{q}^2$.

\begin{figure}
  \centering
  \includegraphics[width=0.5\textwidth]{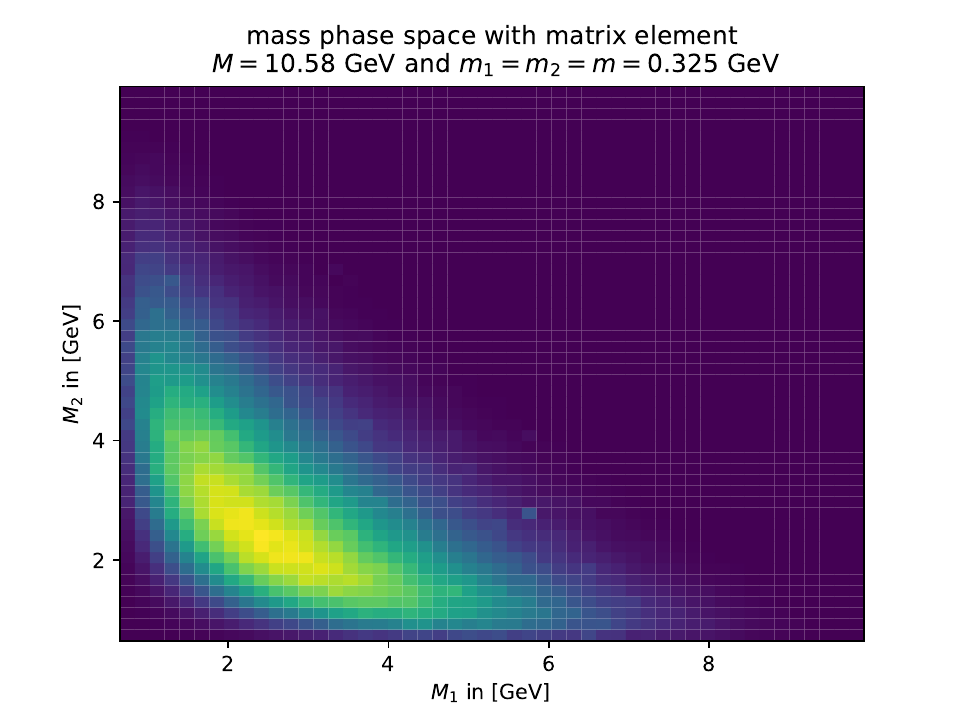}
  \caption{Plot of mass phase space according to the matrix element.}
  \label{fig:MEMassSpace}
\end{figure}

\begin{figure}
  \centering
  \includegraphics[width=0.5\textwidth]{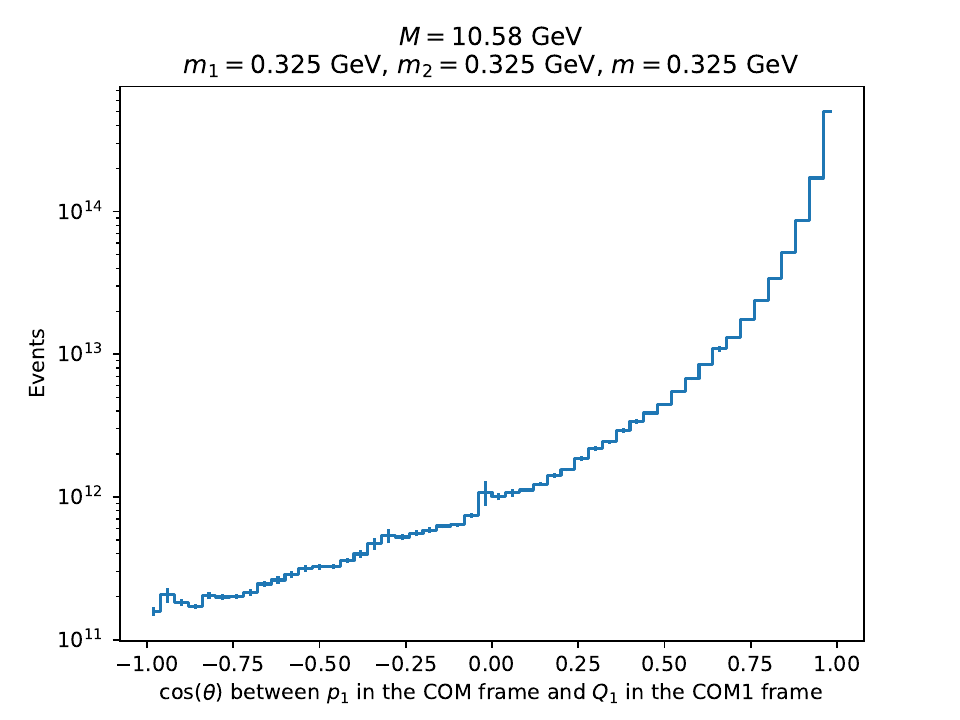}
  \includegraphics[width=0.5\textwidth]{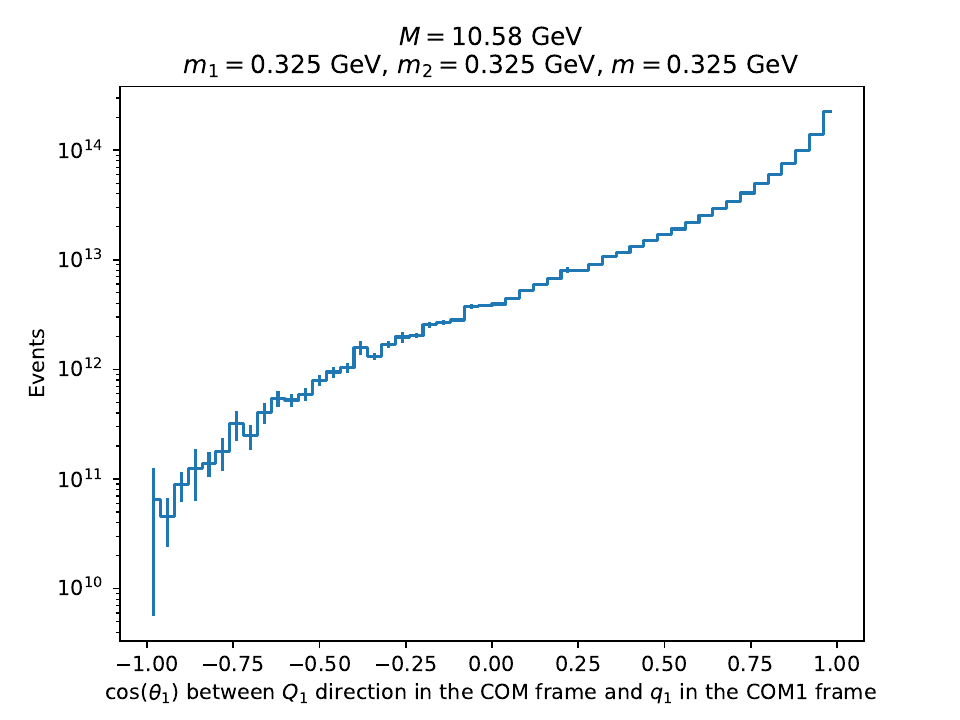}
  \caption{Plot of distributions for angles according to the matrix element.}
  \label{fig:MEAngularDistributions}
\end{figure}

This matrix element does not only have an impact on the directionality
of cluster fission but also a significant contribution to the mass phase
space. In particular the soft function has an explicit dependence on
$M_1,M_2$ through the functions $I_{ij}\propto (M_iM_j+...)^{-2}$. This
means that the matrix element impacts also the mass distributions of the
clusters, which can be seen in Fig.~\ref{fig:MEMassSpace}, where we
generated a weighted histogram of
$\frac{1}{\Gamma_{2\rightarrow 4}}\frac{\measDiff^2 \Gamma_{2\rightarrow 4}}{\measDiff M_1\measDiff M_2}$,
which clearly differs from Fig.~\ref{fig:FlatMassSpace}.
The regulated poles of the soft function drive
the masses down to lower values, while at the same time forbidding the
region, where both clusters are light.
As we want to sample also the angles efficiently, we found their
distributions by using weighted histograms to be quite collinear as
can be seen in Fig.~\ref{fig:MEAngularDistributions}. Note that here
$\theta$ is the angle between $p_1$ and $Q_1$, defined in the original
cluster mass frame (COM) and $\theta_i$ are defined in the final
cluster $Q_i$ rest frame (COM$i$) between the direction of $Q_i$ in
COM and $q_i$. For the normalized distribution
$f_\text{COM}(\cos(\theta))$ we assume the following type of
distribution as a proposal distribution since this is motivated
by the $t$-channel matrix element:
\begin{align}
  f_\text{COM}(\cos(\theta))&=\frac{A(2+A)}{\left[1+A-\cos(\theta)\right]^2}
\end{align}
Here $A>0$ is a parameter of the distribution, which needs to be fitted
to the histogram of the angle, which is plotted in Fig.~
\ref{fig:MEAngularDistributions}. On the other hand for the
$f_\text{COM,i}(\cos(\theta_i))$ proposal distribution we choose an 
exponential ansatz which
is motivated by the almost linear behaviour in the log plot of
Fig.~\ref{fig:MEAngularDistributions}:
\begin{align}
  f_\text{COM,i}(\cos(\theta_i))
  &=\frac{\beta_i}{1-e^{-2 \beta_i }}
    e^{\beta_i (\cos(\theta_i)-1)}
\end{align}
Here we choose $\beta_i>0$ and fit this as well to the histograms at
hand for obtaining the most efficient sampling later on.\par
These parameters $A,\beta_1$ and $\beta_2$ are of course in general
functions of $M,m_1,m_2,m$, which complicates getting good overall
efficiency in the rejection sampling, which we use later on.
Of course if $m_1=m_2$ this implies that $\beta_1=\beta_2$.

\subsection{Implementation of the algorithm in Herwig}

In order to sample from Eq.~\eqref{eq:Gamma24} appropriately and with
unit weights in \code{Herwig} we used the following algorithm for a single
cluster fission:
\begin{enumerate}
	\item \underline{Flavour} --- draw the flavour of the $q\bar{q}$-pair
  using the default weights $\textbf{Pwt}(q)$, which are constant parameters in
  \code{Herwig}. 
\item \underline{Cluster masses} --- sample the masses $M_1,M_2$ uniformly in the
  allowed triangular phase space of Fig.~\ref{fig:FlatMassSpace}
\item \underline{Phase space} ---
  \begin{enumerate}
  \item Boost from the lab frame LAB to the centre of mass frame of the
    original cluster (COM)
  \item Sample the direction of the child clusters according to
    $f_\text{COM}(\cos(\theta))$
  \item Boost in both child cluster's rest frames ($\text{COM}i$ for
    $i\in \{1,2\}$) respectively
  \item Sample the direction of constituents according to
    $f_\text{COM,i}(\cos(\theta_i))$ in the $\text{COM}i$ frames
  \item Boost back to the original cluster rest frame
    $\text{COM}$\label{enum:boost1}
  \item Boost all momenta from to the $\text{COM}$ frame to the $\text{LAB}$ frame \label{enum:boost2}
  \end{enumerate}
\item Accept the configuration with $P_\text{acc}$ 
\end{enumerate}
Note that we sample the masses $M_1,M_2$ uniformly, which - albeit not
being the most efficient choice - is comparatively fast if the actual
integrated mass distribution
$\frac{1}{\Gamma_{2\rightarrow 4}}\frac{\measDiff^2 \Gamma_{2\rightarrow
    4}}{\measDiff M_1\measDiff M_2}$ is not known. Furthermore we always sample the \linebreak
azimuthal angles $\phi_\text{COM},\phi_\text{COM,1}$ and
$\phi_\text{COM,2}$ uniformly \linebreak from $[0,2\pi]$.  Note that the two boosts
in the points \ref{enum:boost1} - \ref{enum:boost2} are necessary since
we cannot simply boost from the $\text{COM}i$ frames to the $\text{LAB}$
frame.  Here, we would pick up a rotation since non-parallel Lorentz
boosts do not commute. In
the algorithm above we have to still define $P_\text{acc}$, which we
define as the following:
\begin{align}
  P_\text{acc}=\frac{f_\text{PS}(M_1,M_2)
  |\mathcal{M}|_{2\rightarrow 4}^2
  (\{p_i\cdot p_j\})}{\lambda_\text{OE}f_\text{MS}f_\text{COM}
  (c_{\theta})f_\text{COM,1}( c_{\theta_1})f_\text{COM,2}(c_{\theta_2})}
\end{align}
Here we abbreviated $\cos(\theta_i)\equiv c_{\theta_i}$,
$\{p_i\cdot p_j\}$ stands for the minimal set of all possible scalar
products of the sampled set of four-momenta and $\lambda_\text{OE}$ is
the overestimate of the expression such that ideally
$P_\text{acc}\leq 1$ for all phase space. In addition, $f_\text{MS}$
could be a pre-sampler of the masses $M_1,M_2$, but since we sample the
masses uniformly we set $f_\text{MS}=1$ without loss of generality. One
should note that generally the best overestimate $\lambda_\text{OE}$ is a
function of $M,m_1,m_2$ and $m$. Furthermore, finding such an
overestimate is highly non-trivial especially due to the denominator
probability density functions. We therefore choose to get an
estimate of the overestimate, by sampling the phase space and searching
for a maximum of $\lambda_\text{OE}P_\text{acc}$ for fixed $m_1,m_2$ and
$m$ but variable $M$.

\subsection{Phase space and matrix elements for cluster decay}
\label{sec:NewCD}

The current cluster decay model of \code{Herwig} \cite{Bewick:2023tfi} forces
clusters to decay to two hadrons, where the selection of a given hadron
pair depends on different weight factors.  These include a multiplicity
factor for angular momenta, flavour multiplet weights for the individual
hadrons and, most importantly a weight for the available phase space
volume, which is given by the momentum of the decay products in the
cluster rest frame, such that heavier hadrons are suppressed due to smaller
momentum that would be available.  Note, that baryons are treated
separately from mesons, similar to the model in \cite{Kupco:1998fx}.  

If a cluster cannot decay into two hadrons because its mass is too
small to split into two hadrons a reshuffling with nearby clusters is
performed and the cluster will be decayed to a single hadron. The
kinematics of the decay are depending on whether at least one of the
cluster's constituents have originated from the hard process or the
parton shower and have a virtuality $q>m_{g}$, where $m_{g}=0.95$\,GeV
is the gluon constituent mass, or the cluster has been produced during
the cluster fission. In the latter case the cluster is decayed fully
isotropically in its rest frame. In the first case the hadron
directions $c_\theta=\cos (\theta_{p_1,h_1})$ (defined in the COM
frame of the cluster with respect to the initial constituent partons)
are sampled according to the distribution
\begin{align}
	\label{eq:DefaultCD}
  f\left(c_{\theta}\right) = \frac{1}{\lambda} e^{\frac{c_{\theta}-1}{\lambda}} \ .
\end{align}
This picture of treating clusters in a two-fold manner is contrary to
the consequences from the colour evolution picture that the cluster
fission should be treated perturbatively and the shower cutoff should be
considered merely as a factorization scale \cite{Platzer:2022jny}.

\begin{figure}[t]
  \centering
  \includegraphics[width=0.25\textwidth]{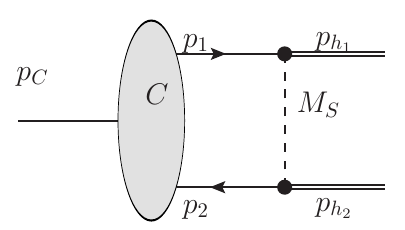}
  \caption{Diagram of the chosen cluster decay matrix
    element. Currently we use this to sample the directions of the
    outgoing hadrons. In particular thresholds and the full mass
    dependence will be added later.}
  \label{fig:FeynmanDiagramCD}
\end{figure}

We therefore present a new cluster decay model that tries to
interpolate the cluster fission kinematics and the cluster decay
kinematics in a more meaningful way. Similarly to the cluster fission,
we assume a matrix element $\mathcal{M}_{p_1,p_2\rightarrow h_1,h_2}$,
which in this case is a $2\rightarrow 2$ parton to hadron process.  We
choose a $t$-channel like interaction of the following form as
displayed in Fig.~ \ref{fig:FeynmanDiagramCD}:
\begin{align}
  |\mathcal{M}_{p_1,p_2\rightarrow h_1,h_2}|^2
  =& \frac{A_0}{[(p_1-h_1)^2-M_S^2]^2} \nonumber\\
  =& \frac{A_0'}{[A-\cos(\theta_{p_1,h_1})]^2} \nonumber
\end{align}
Here $M_S$ is a pseudo mass parameter that we set to
$M_S=\max\{(m_1-m_{h_1}),(m_2-m_{h_2})\}$ in order to have the maximally
aligned cluster decay for small momentum constituents\footnote{Note that
  the maximum has to be taken since in general
  $p_1\cdot h_1>\max\{m_1 m_{h_1},m_2 m_{h_2}\}$}. $A_0, A_0'$ are
arbitrary normalization constants.

While still a model assumption, the most important difference to the
previous model obviously is the choice of kinematics and the
$t$-channel behaviour.  This will ensure a smooth continuation of the
kinematics from the parton level into the hadron level.  We will
demonstrate the consequences of this choice below.

\section{Impact on existing observables}
\label{sec:Benchmarks}

Within the steps towards a new, evolving, hadronization model we are
now mainly interested in finding observables which are able to help
constrain the individual building blocks and where we can already see
improvements from the modifications we propose.  We will start by
considering existing observables, and we limit ourselves to
$e^+e^-\rightarrow q\bar{q}$ collision data such that multiple
partonic interaction effects are absent and colour reconnection
effects are not overpowering the cluster fission and decay effects
that we want to study in this paper.
\subsection{LEP energies}
\begin{figure}
  \centering
  \includegraphics[width=0.5\textwidth]{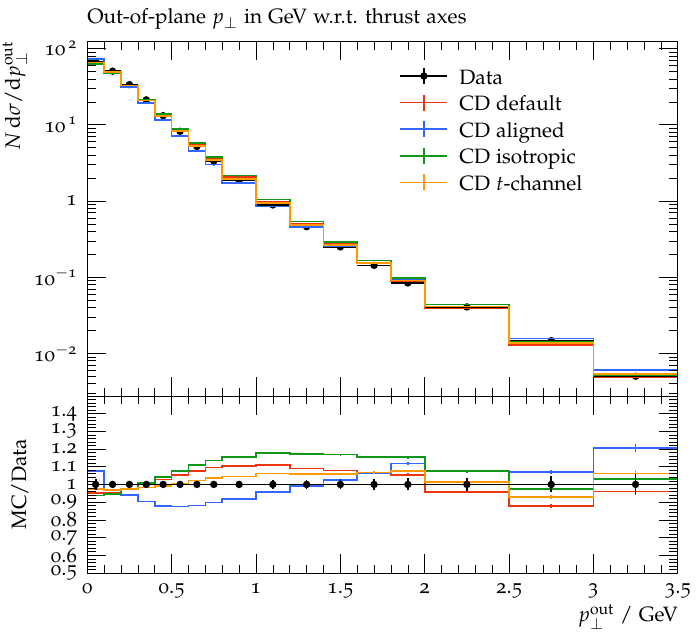}
  \includegraphics[width=0.5\textwidth]{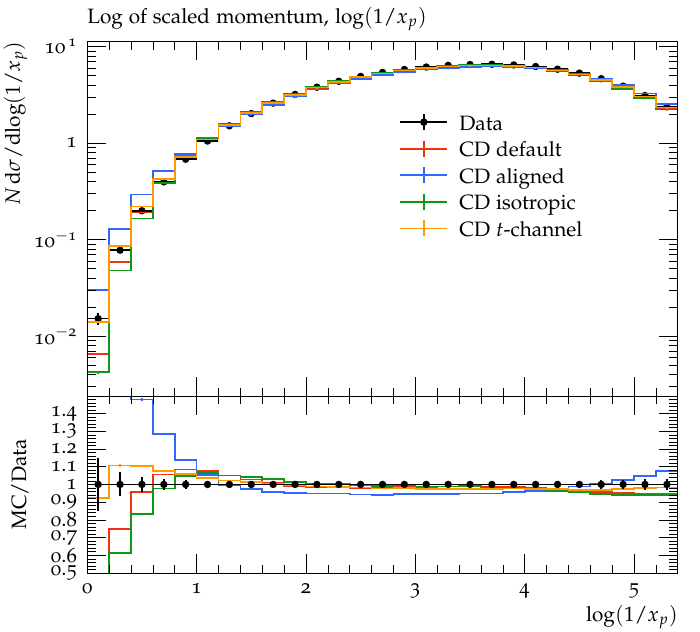}%
  \caption{The $p_T^{\text{out}}$ distribution and the
    $\zeta_p=-\log x_p$ distribution measured by DELPHI
		\cite{DELPHI_1996_S3430090}, where $x_p=2|\vec{p}|/\sqrt{s}$. We show a variation of different
    cluster decay models (keeping the \code{Herwig} default cluster fission),
    where `default' is the \code{Herwig} default, `aligned' is a fully aligned
    CD, `isotropic' is a completely isotropic CD and $t$-channel is our
    matrix element ansatz for the CD.}
  \label{fig:LEP_pTin}
\end{figure}
An obvious first test of the model is a comparison with data from LEP
where most of the current hadronization models are tuned against
particle yields and event shape data. In order to compare to experimental data
we use the public analysis framework \code{Rivet} \cite{Rivet}.

In Fig.~\ref{fig:LEP_pTin} we show the distribution of transverse momenta
relative to the thrust axis and the scaled momentum distribution at LEP
for different cluster decay models and compare them to the `default' \code{Herwig}
model. Two of these choices are somewhat extreme.
In the `aligned' scheme the two hadrons move fully aligned with the original quarks.
In the `isotropic' scheme, the decay is fully isotropic which resembles the 
`default' choice with the exception that also heavy quark containing clusters 
are fissioned completely isotropically.

Both observables show some sensitivity to the cluster decay model,
especially for the low and high momentum regions. Nonetheless, we
found that the new `$t$-channel' scheme -- using the ansatz for a
matrix element with a $t$-channel propagator as described in section
\ref{sec:NewCD} -- does not introduce new tensions in the LEP
observables. In fact, e.g.\ for the observables in
Fig.~\ref{fig:LEP_pTin} improvements with respect to the default model
can be seen, where we should note that we left all other parameters at
the tuned default values in order to highlight the differences between
the model choices.

\subsection{Low energy $e^+e^-$ data}

Much more information will be available from data taken at lower
energies, {\it e.g.} at B-factories, where the effect of the parton
shower is small and in our picture the clusters are often formed
directly by an initial $q\bar q$ system that will form a single
primordial cluster that can decay, after one or at most very few
possible fissions, into hadrons.  In contrast, already at LEP the
partons form jets and the properties of individual particle production
only has a small effect on the properties of jet observables or event
shapes.

We show some results of the newly proposed cluster fission algorithm
for the di-hadron fragmentation function as a function of
$m_{h_1h_2}$, binned in $z=2(E_{h_1}+E_{h_2})/\sqrt{s}_\text{BELLE}$
measured by BELLE \cite{BELLEanalysis}, where $m_{h_1h_2}$ is the invariant mass of two
hadrons within the same event in the same hemisphere\footnote{with respect to the thrust direction.}
and $E_{h_i}$ is 
their respective energy. This observable is very sensitive to cluster
fission for $z\rightarrow 1$ since in this case the parton shower is 
essentially not able to emit any partons above the parton shower cutoff and
the particle production comes mostly from hadronization. \linebreak Therefore, this
region is dominated mostly by the recursive cluster fission of a
$M_C\lesssim \sqrt{s}_\text{BELLE}$ mass cluster.

\begin{figure}[t]
  \centering
  \includegraphics[width=0.5\textwidth]{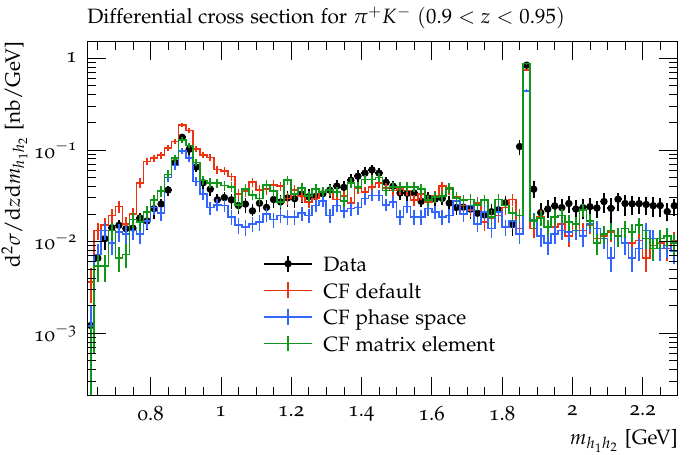}
  \includegraphics[width=0.5\textwidth]{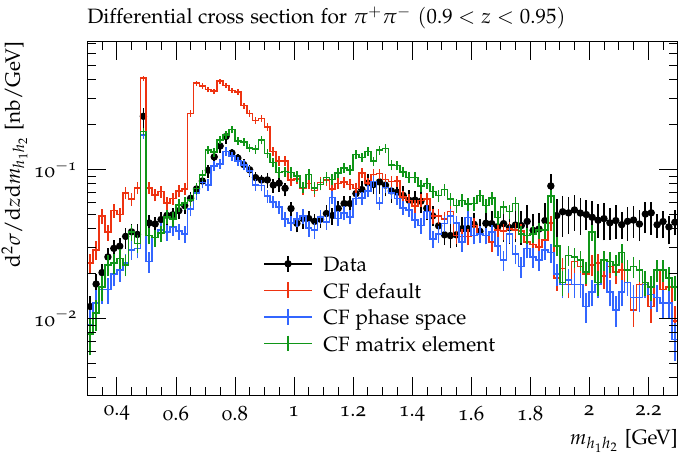}%
  \caption{Comparison of the different cluster fission models with
    respect to BELLE data from \cite{BELLEanalysis}. Here the $t$-channel like
  cluster decay was used.}
	\label{fig:BELLE_dihadron_CFvar}
\end{figure}

\begin{figure}[t]
  \centering
  \includegraphics[width=0.5\textwidth]{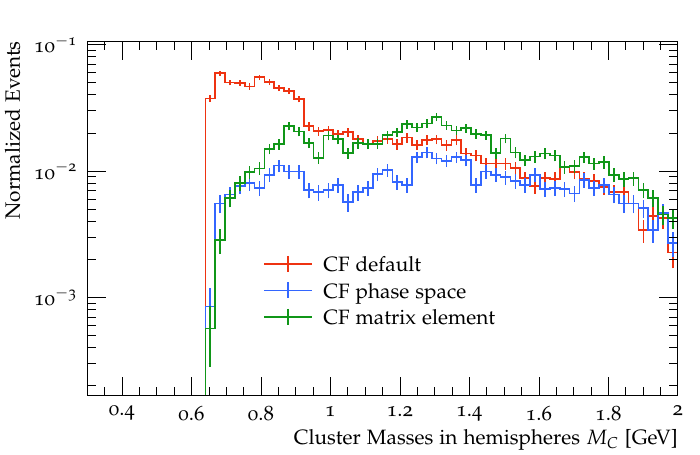}%
  \caption{The mass distribution of clusters binned in $\tilde{z} \in [
    0.9 , 0.95 ]$. Here the $t$-channel like cluster decay was used.}
  \label{fig:HerwigClusterMasses}
\end{figure}

In Fig.~\ref{fig:BELLE_dihadron_CFvar} we compare the following three
options for the cluster fission model, where for all of these the new 
$t$-channel like cluster decay model has been used:
\begin{enumerate}
\item \code{Herwig} 7.3.0.\ default fully aligned cluster fission (red)
\item The new model with the flat mass phase space and fully aligned fission
  (blue) according to Eq.~\eqref{eq:MassPhaseSpace}
\item The new model with the full matrix element dependence (green) and
  underlying mass distributions as shown in Fig.~\ref{fig:MEMassSpace}
\end{enumerate}
We can clearly see that the \code{Herwig} `default' forms a \linebreak plateau
starting from $(m_{q_1}+m_{q_2})$ until $\sim 0.9$ GeV, which can be
explained by the fact that for $\textbf{PSplit}\sim 1.0$ the cluster
mass distribution is a triangular distribution as also shown in
\cite{Hoang:2024zwl}. This plateau is not reflected in the data, which hints
at a breakdown of the default model for cluster fission. On the other
hand we can see that the flat phase space and the full matrix element
yield substantially better results except for the mid and high
$m_{h_1h_2}$ regions, where for the tail region none of the models
describes the data accurately. The discrepancy in the mid region can be
attributed at least partly to the fact that the new models are still
untuned.

In Fig.~\ref{fig:HerwigClusterMasses} we plot the unphysical (but informative)
cluster mass distribution $f(M_C)$ after the recursive cluster fission binned
in $\tilde{z}=2E_C/\sqrt{s}_\text{BELLE}$, where $E_C=E_1+E_2$ is the
energy of the cluster that decays into two hadrons of energies $E_1$ and $E_2$.
Looking at the cluster mass distributions one can clearly see that indeed the
new cluster fission models are solving the plateau issue. In fact one can see
that the hard turn-on of the cluster masses is regulated by the constituent 
masses of the quarks and the threshold functions of the phase space yielding 
to a smooth turn-on. This regularization of the infra-red divergence by
massive partons is also a well-known feature of the parton shower.

\begin{figure}[t]
	\centering
	\includegraphics[width=0.5\textwidth]{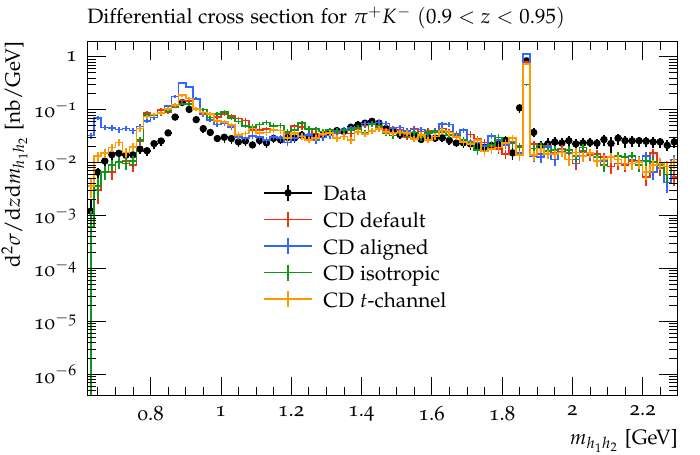}
	\includegraphics[width=0.5\textwidth]{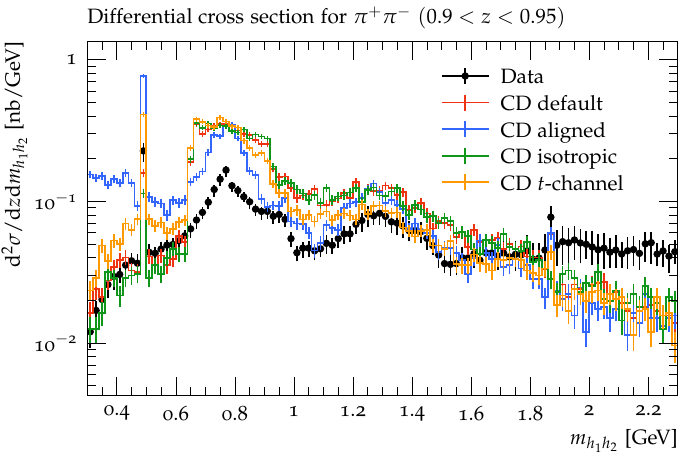}%
	\caption{Comparison of the different cluster decay models with
		respect to BELLE data from \cite{BELLEanalysis}. We used
		the default cluster fission model.}
	\label{fig:BELLE_dihadron_CDvar}
\end{figure}

We also examined the variation of the cluster decay model on these BELLE observables
in Fig.~\ref{fig:BELLE_dihadron_CDvar}, but with the exception of the extreme cases 
the impact of the cluster decay is rather limited.

\section{Dissecting the hadronization history\\ with new observables}
\label{sec:NewObservables}

As we have seen in the previous section some unphysical features could
be successfully removed qualitatively, while not producing different
unphysical artefacts in other well established data. Nonetheless we
want to study more in depth to which observables the cluster fission
and the cluster decay model are most sensitive. To this end we present
in this section phenomenological studies of the generalized jet
angularities and the multi point energy correlators. Since we want to
find classes of observables that can be sensitive to both cluster
fission and cluster decay we show in the following only observables at
$\sqrt{s}=M_Z$ unless stated otherwise.
\subsection{Angularities}
\label{sec:angularities}
\begin{figure}
	\centering
	\includegraphics[width=0.5\textwidth]{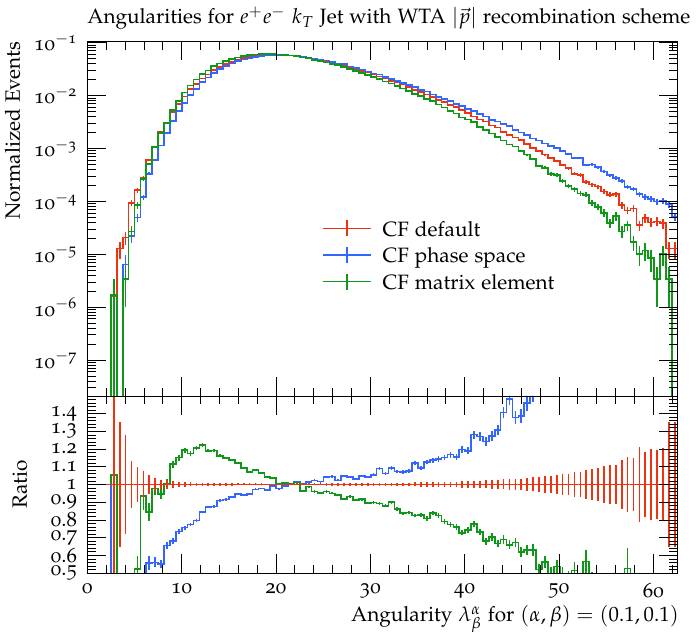}
	\includegraphics[width=0.5\textwidth]{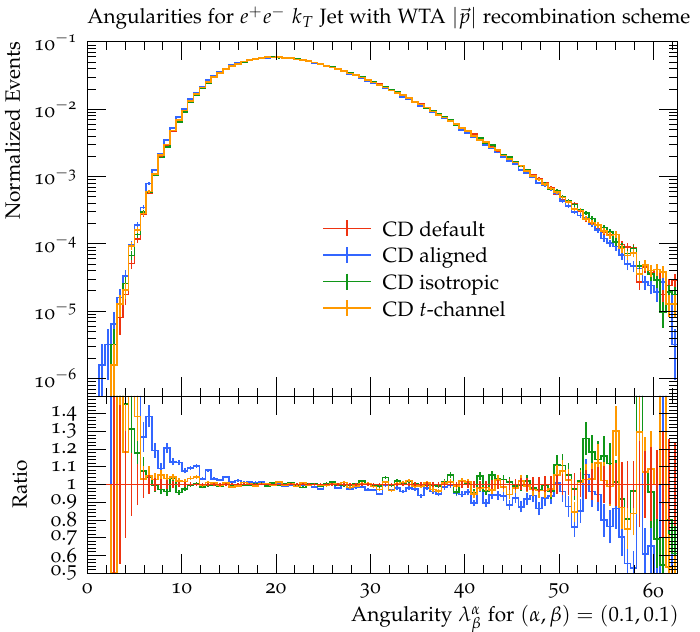}%
	\caption{The generalized jet angularities for $\alpha =
          0.1,\beta=0.1$ for the $k_t$ WTA algorithm. We show a
          variation of the cluster fission models with the $t$-channel
          like cluster decayer (top) and the different cluster decay
          models with the \code{Herwig} default cluster fission
          (bottom).}
	\label{fig:AngularitiesCDunsensCFsens}
\end{figure}
\begin{figure}
	\centering
	\includegraphics[width=0.5\textwidth]{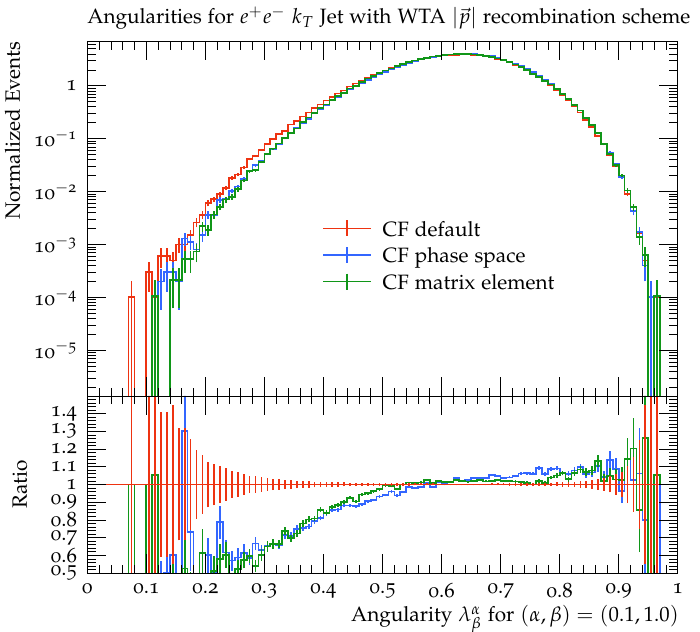}
	\includegraphics[width=0.5\textwidth]{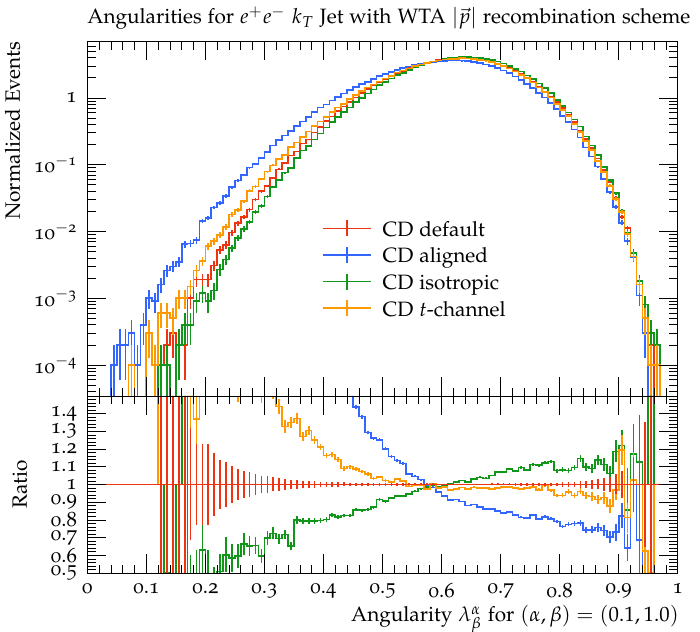}%
	\caption{The generalized jet angularities for $\alpha =
          0.1,\beta=1.0$ for the $k_T$ WTA algorithm. We show a
          variation of the cluster fission models with the $t$-channel
          like cluster decayer (top) and the different cluster decay
          models with the \code{Herwig} default cluster fission
          (bottom).}
	\label{fig:AngularitiesCDsensCFunsens}
\end{figure}

In addition to the improved behaviour for the measured observables in
Fig.~\ref{fig:LEP_pTin} and Fig.~\ref{fig:BELLE_dihadron_CFvar} we also
show generalized angularities and multi-energy correlations, which \linebreak are
sensitive to the different stages of hadronization. The generalized
angularities $\lambda_\beta^\alpha$ introduced in 
\cite{Angularities,Angularities2} are defined as
\begin{align}
  \label{eq:Angularities}
  \lambda_\beta^\alpha = \sum_{i\in \{\text{jet}\}}z_i^\beta
  \theta_i^\alpha \ ,
\end{align}
where $\theta_i:=\sqrt{2(1-\cos(\hat{\theta}_i))}$ is defined with
$\hat{\theta}_i$ being the angle between jet constituent $i$ and the
jet. The energy fraction $z_i=\frac{E_i}{E_\text{vis}}$ is defined as
$E_\text{vis}=\sum_{i\in \{\text{event}\}}E_i$ and gives a measure of
hardness. Different powers of $\beta$ and $\alpha$ provide different
weighting of softness and collinearity respectively and give insights
into different regimes of the infra-red and collinear limit. Note in particular
that the angularities $\lambda^\alpha_\beta$ are infra-red safe for $\beta>0$
and additionally collinear safe if $\beta=1$ and $\alpha \geq 0$.
As a jet algorithm we used the  generalized $k_T$ algorithm with the
Winner-Takes-All (WTA) $|\vec{p}|$-recombination scheme
\cite{WTAscheme,WTAscheme2} from \code{FastJet} \cite{FastJet} to cluster 
each event into $2$ exclusive jets.

In Fig.~\ref{fig:AngularitiesCDunsensCFsens} and
Fig.~\ref{fig:AngularitiesCDsensCFunsens} we show various generalized
angularities for the case of $e^+e^-$ collisions at $\sqrt{s}=M_Z$.
In particular we compare different cluster fission models on top and
different cluster decay models on the bottom for $\alpha=0.1$ and
different $\beta$. We can clearly see that in
Fig.~\ref{fig:AngularitiesCDunsensCFsens} the cluster decay model is
irrelevant, but the cluster fission model has a strong impact on the
observable. This can be attributed to the fact that in the limit
$\alpha,\beta \rightarrow 0$ the angularities become the jet
multiplicity and since only the cluster fission can produce
multiplicity, the cluster decay is irrelevant to this observable. In
Fig.~\ref{fig:AngularitiesCDsensCFunsens} we see the opposite
behaviour i.e.\ the cluster decay has a strong impact on the
observable, while the cluster fission has only limited impact on the
observable.

\begin{figure}
  \centering
  \includegraphics[width=0.5\textwidth]{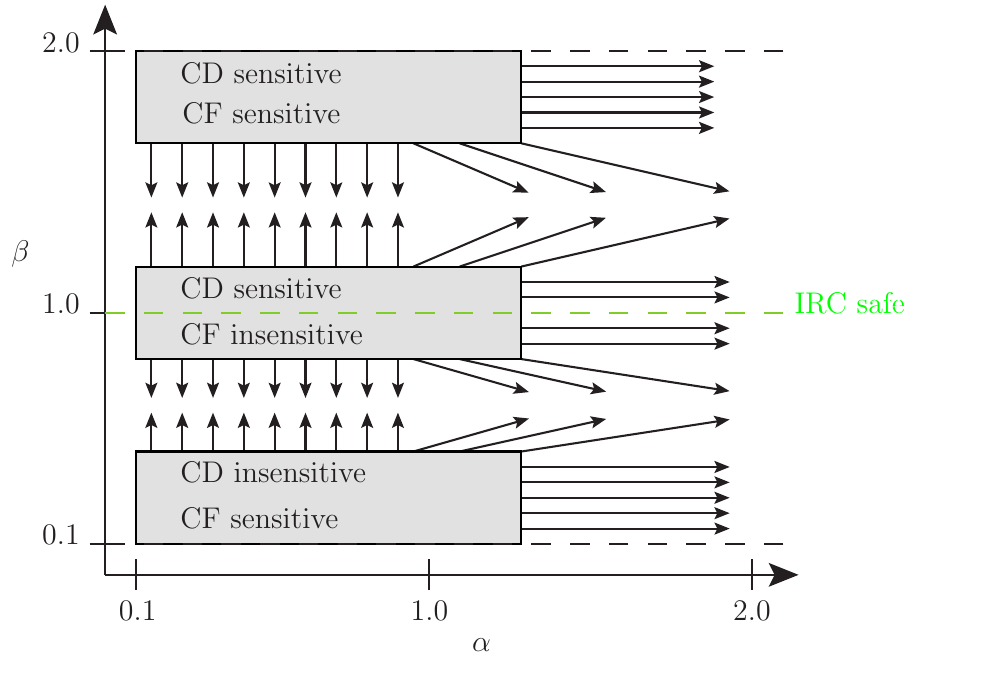}
  \caption{Summary sketch of the angularity parameter space of
    $\alpha,\beta$ with its corresponding sensitivities to the cluster
    fission and cluster decay. The arrows point towards decreasing
    sensitivities and the boxes represent the regions with the most
    sensitivity and discriminative power.}
	\label{fig:AngularitiesSensitivties}
\end{figure}

We have studied several different values in the $(\alpha,\beta)$
space and summarize our findings in the sketch
Fig.~\ref{fig:AngularitiesSensitivties}, where we show the regions
with the most sensitivity as grey blocks and the arrows point towards
decreasing sensitivity in general. It is remarkable that the cluster
fission model shows the least sensitivity in
Fig.~\ref{fig:AngularitiesCDsensCFunsens}, i.e.\ for the collinear
safe angularities, while the cluster decay has a huge impact on this
observable.  Note that this observable in the limit
$\alpha\rightarrow 0$ is just the energy fraction of the jet.

\subsection{Correlations}
\label{sec:corrs}
In addition to the angularities we looked at weighted multi-point energy
correlations $\text{ENC}_\gamma(\theta)$, which are defined as 
\begin{align}
  \label{eq:ENCs}
  \text{ENC}_\gamma (\theta)
  &=\frac{1}{\sigma_\text{tot}}\sum_{m=N}^\infty \int \text{d}\sigma_m
  \mathcal{W}_{m}^{\gamma}[\theta,\{p_{i}\}_m]\\ \text{d}\sigma_m&=\text{d}\sigma({e^+e^-\rightarrow
    p_1,...,p_m})\\ \mathcal{W}_{m}^\gamma[\theta,\{p_{i}\}_m]&=\frac{1}{D_{m}^{\gamma}}\sum_{i_1<...<i_N}^m
  \left(\prod_{k=1}^N
  E_{i_k}^\gamma\right)\delta(\hat{\theta}(\{\hat{p}_{i_k}\}_{N})-\theta) \label{eq:EnergyPowers}\\ D_{m}^{\gamma}&=\sum_{i_1<...<i_N}^m\left(\prod_{k=1}^N
  E_{i_k}^\gamma \right)\ .
\end{align}
Here, $\hat{\theta}(\{\hat{p}_i\}_N)$ is a symmetric function of the $N$
particle directions on the subset $\{p_i\}_N\subseteq \{p_i\}_m$ of the
$m$ particle final state $\{p_i\}_m$. Note that the $N$-point
correlations $\text{ENC}_\gamma(\theta)$ are normalized to unity, with
the total inclusive cross section
$\sigma_\text{tot}\equiv\sum_{m=2}^\infty\sigma_m$.  Due to the
computational cost, which is $\mathcal{O}(m^N)$, of the
$\text{ENC}_\gamma(\theta)$ for large $m$ we limit ourselves to the
cases $N\in\{2,3\}$. For $N=2$ the regular infrared safe energy-energy 
correlations are recovered if the function
$\hat{\theta}(\hat{p}_1,\hat{p}_2)\equiv \theta_{12}$ and $\gamma=1$ is
chosen, but since we are not reliant on any infrared safety condition
the value of $\gamma$ can be an arbitrary real number and in particular
also negative values, which would lead to an \textit{infrared dangerous}
observable. In this study we looked at the values
$\gamma \in \{0,\pm 1, \pm 2,\pm \infty\}$, even though fractional
values are also possible.  We call $\gamma\rightarrow +\infty$ the
`Winner-Takes-All' (WTA) correlation and $\gamma\rightarrow -\infty$ the
`Soft-Takes-All' (STA) correlation, which corresponds to filling only
the largest product of energies per event for WTA (smallest for STA).

\begin{figure}
	\centering
	\includegraphics[width=0.5\textwidth]{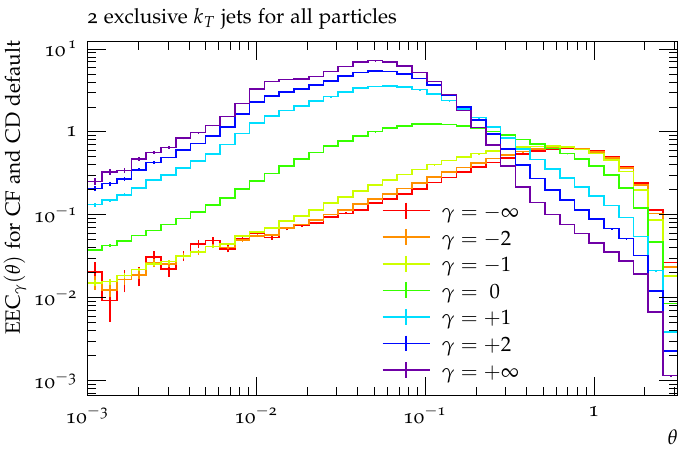}
  \caption{We show the EEC$(\theta)$ correlation of two exclusive $k_T$ jets
  for variable $\gamma$ with the default cluster fission and cluster decay
model.}
	\label{fig:LEP_Correlations_GammaInterpolation}
\end{figure}

In practise we find that the correlations for any finite $\gamma$ we studied
are qualitatively similar to an interpolation between the WTA and STA
correlations as we show e.g.\ in
Fig.~\ref{fig:LEP_Correlations_GammaInterpolation} for the observable
EEC$(\theta)$. Therefore, we only consider the WTA and STA correlations as most
extreme cases in the following.

While for $N=2$ we choose the relative angle $\theta_{ij}$ between the two
particles in the lab frame, for $N=3$ we choose the function
$\hat{\theta}(\hat{p}_1,\hat{p}_2,\hat{p}_3)$ as the minimum and maximum
of the pair-wise relative angles $\theta_{ij}$:
\begin{align}
  \label{eq:thetaMin}
  \hat{\theta}_\text{min}(\hat{p}_1,\hat{p}_2,\hat{p}_3)=\min_{i<j} \{\theta_{ij}\} \\
  \label{eq:thetaMax}
  \hat{\theta}_\text{max}(\hat{p}_1,\hat{p}_2,\hat{p}_3)=\max_{i<j} \{\theta_{ij}\}
\end{align}
Note that neither the choice of $\hat{\theta}_\text{min}$ or $\hat{\theta}_\text{max}$
is a collinear safe choice since we let the sum in Eq.~\eqref{eq:EnergyPowers} run over 
distinct ordered tuples of particles instead of all possible tuples (compare e.g.\ \cite{ENC_Massimiliano}). However in 
the quest for studying soft physics we are not limited to collinear safe observables either.

\begin{figure}
	\centering
	\includegraphics[width=0.5\textwidth]{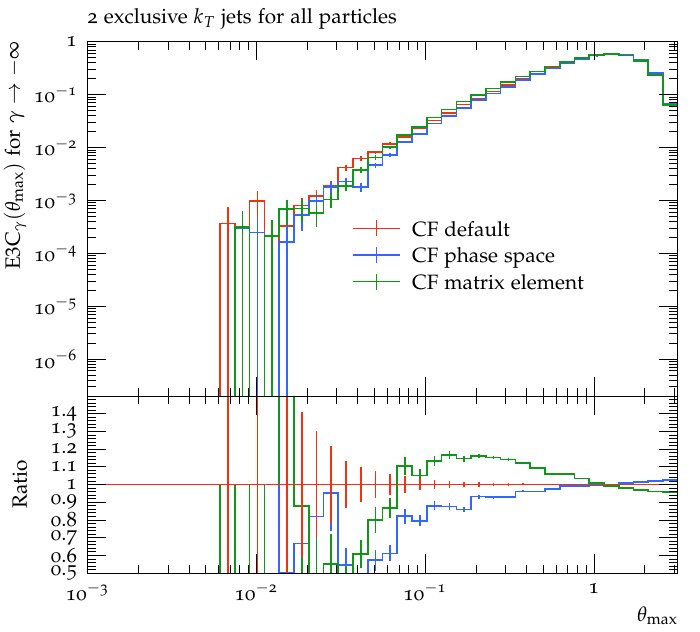}
	\includegraphics[width=0.5\textwidth]{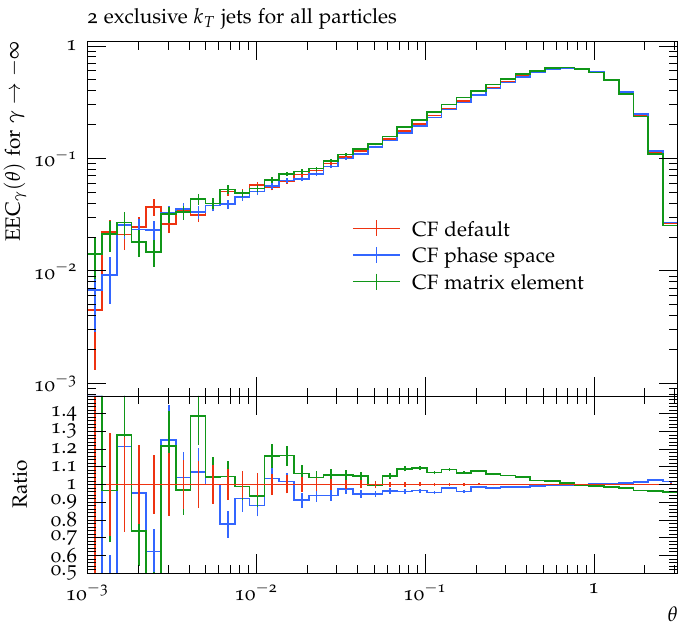}
	\caption{The STA ($\gamma\rightarrow - \infty$) energy
          correlator E$3$C$_\gamma(\theta_\text{max})$ (top) and
          EEC$_\gamma(\theta)$ (bottom) of two exclusive $k_T$
          jets. We show a variation of the cluster fission models with
          the $t$-channel like cluster decayer.}
	\label{fig:LEP_Correlations_EEC_E3C_thetaMax_STA}
\end{figure}

In addition to the inclusive final state energy correlations we looked
at energy correlations within two exclusive jets clustered using the
Durham $k_T$ algorithm \cite{DurhamKT} with the $E$ recombination scheme. For this we
computed the $\text{ENC}_\gamma$ on the set of particles within each jet
and scaled the overall observable by the number of jets, here two, such
that the normalization to unity is recovered.

In Fig.~\ref{fig:LEP_Correlations_EEC_E3C_thetaMax_STA} we display the
STA $\text{E3C}_\gamma(\theta_\text{max})$ and
$\text{EEC}_\gamma(\theta)$ for a variation of the cluster fission model
with the $t$-channel like cluster decayer. We note that the observable
is sensitive to the choice of the cluster fission model. The same
observables were studied for fixed cluster fission and variable cluster
decay model, which showed that they were not at all dependent on the
cluster decay. In fact, all the STA energy correlations we studied were
essentially oblivious to the cluster decay model.  This observation
implies that these STA correlations are able to discriminate between
different cluster fission models but at the same time they are not
affected by the cluster decay model.

\begin{figure}
	\centering
	\includegraphics[width=0.5\textwidth]{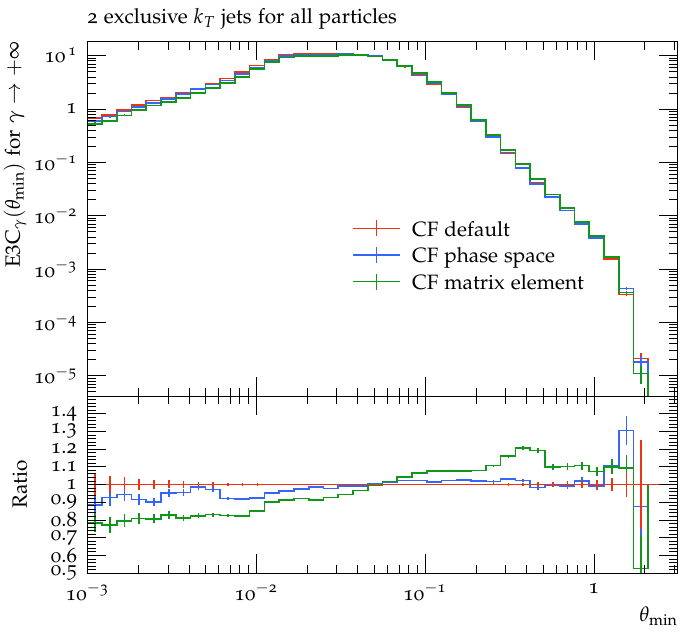}
	\includegraphics[width=0.5\textwidth]{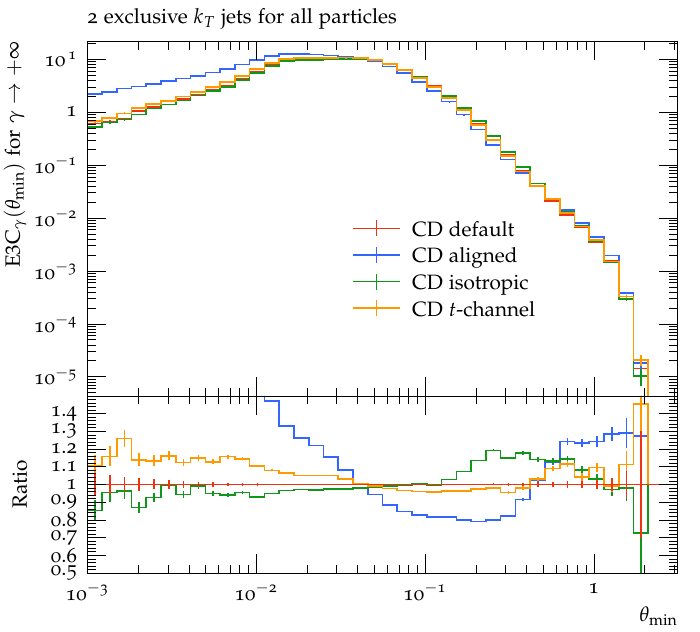}%
	\caption{The WTA ($\gamma\rightarrow + \infty$) energy
          correlator E$3$C$_\gamma(\theta_\text{min})$ of two
          exclusive $k_T$ jets. We show a variation of the cluster
          fission models with the $t$-channel like cluster decayer
          (top) and the different cluster decay models with the
          \code{Herwig} default cluster fission (bottom).}
	\label{fig:LEP_Correlations_E3C_thetaMin_WTA}
\end{figure}

\begin{figure}
	\centering
	\includegraphics[width=0.5\textwidth]{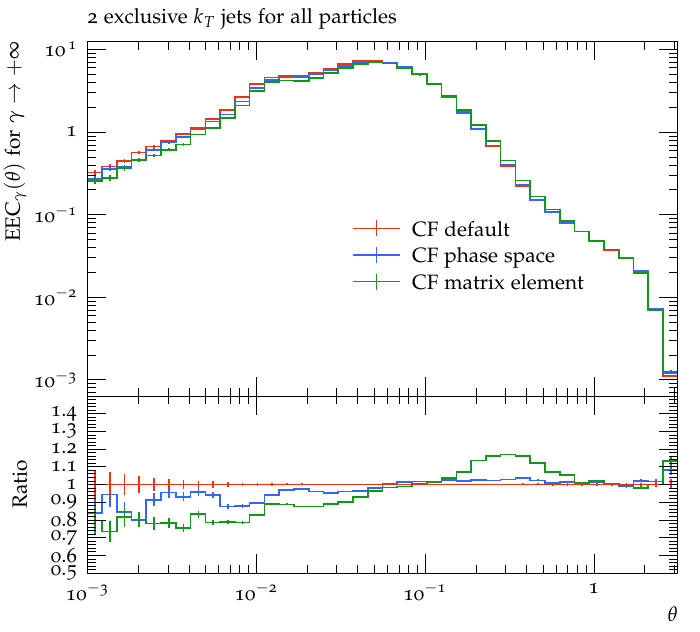}
	\includegraphics[width=0.5\textwidth]{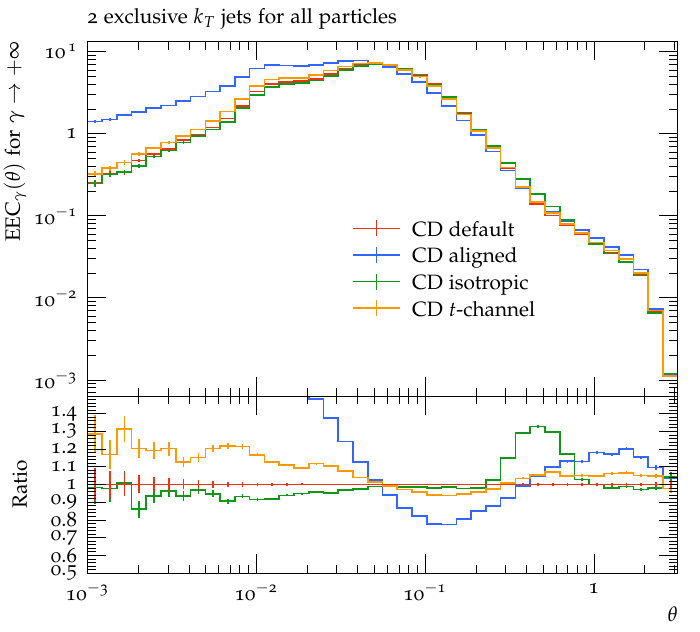}%
	\caption{The WTA ($\gamma\rightarrow + \infty$) energy
          correlator EEC$_\gamma(\theta)$ of two exclusive $k_T$
          jets. We show a variation of the cluster fission models with
          the $t$-channel like cluster decayer (top) and the different
          cluster decay models with the \code{Herwig} default cluster
          fission (bottom).}
	\label{fig:LEP_Correlations_EEC_WTA}
\end{figure}

In contrast, for the WTA $\text{E$3$C}_\gamma(\theta_\text{min})$ correlations, \linebreak
which we show in Fig.~\ref{fig:LEP_Correlations_E3C_thetaMin_WTA}, sensitivity to both cluster 
fission (top) and cluster decay (bottom) can be observed. 
In Fig.~\ref{fig:LEP_Correlations_EEC_WTA} we show the
WTA $\text{EEC}(\theta)$ correlations for different
cluster fission models (top) and different cluster decay models
(bottom). As one can clearly see, both are sensitive to this observable,
but the cluster fission has almost no impact at large angles $\theta>1$,
while the cluster decay is significantly sensitive. Note that the
smaller secondary bump at $\theta\simeq 10^{-2}$ comes from the hadron
decay model.

\begin{figure}
  \centering
  \includegraphics[width=0.5\textwidth]{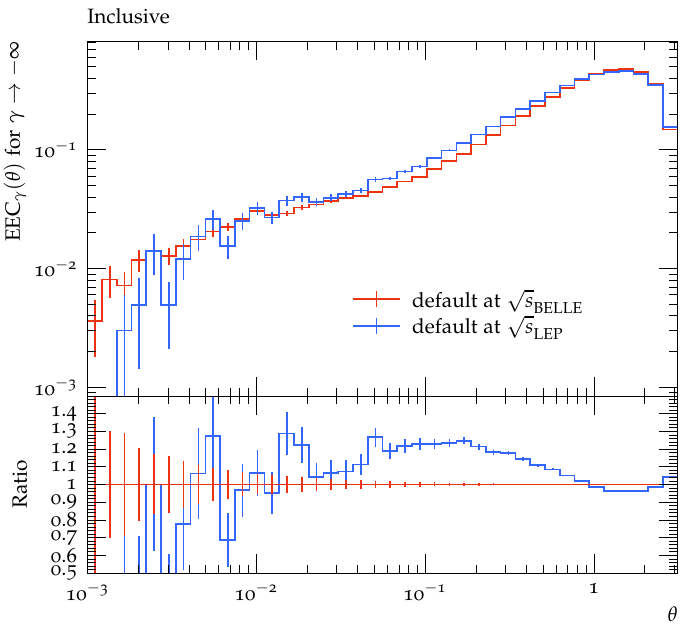}
  \includegraphics[width=0.5\textwidth]{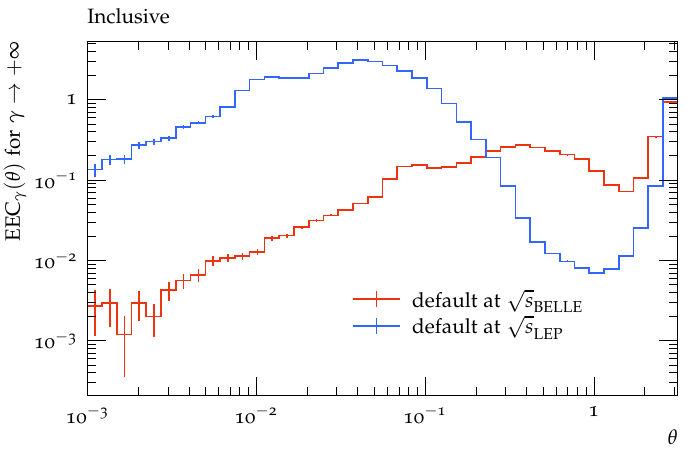}%
  \caption{The STA (top) and WTA (bottom) energy correlators EEC$_\gamma(\theta)$ inclusive over
    the whole event for $\sqrt{s}\in \{10.58\text{ GeV},91.2\text{
      GeV}\}$. Both plots are made with the \code{Herwig} default
    cluster fission and decay.}
	\label{fig:ENERGYvar_Correlations}
\end{figure}

To assess the dependence of these correlations on different centre of
mass energies $\sqrt{s}$, we compare the STA and WTA energy correlations
EEC$_\gamma (\theta)$ for two different
$\sqrt{s}\in \{\sqrt{s}_\text{BELLE},M_Z\}$ in
Fig.~\ref{fig:ENERGYvar_Correlations}. One can clearly see that for the
STA correlations the distributions look quite similar, as one would expect, 
since this observable is infra-red dangerous, i.e.\ the addition of a
hard particle to the event does not change\footnote{In principle, an
  infinitely hard particle, but one can rephrase this as rescaling all
  other particles to become infinitely soft.} the observable. In other
words, this observable measures the soft structure of a fragmenting
unbound $q\bar{q}$ state produced at energies $\sqrt{s}$, which is
independent of the hard scale $\sqrt{s}$ if hadronization universality
is assumed. The residual difference between the two lines can be partly
attributed to the hadronization power corrections of
$\mathcal{O}(\Lambda_\text{QCD}/\sqrt{s})^p$.

On the other hand we show the WTA correlations in
Fig.~\ref{fig:ENERGYvar_Correlations}.  In this case of course we
expect large differences since the observable is IR safe and thus very
sensitive to the hard scale $\sqrt{s}$. One can see that the collinear
peak shifts to lower angles for higher energies.

\section{Conclusion and Outlook}
\label{sec:Conclusion}

In the present study we have started to build a novel structure for
the cluster hadronization model, which is driven by low-scale
perturbative processes driving the cluster fission and eventually
non-perturbative cluster decays. We have in particular focused on
determining observables \linebreak which will be sensitive to such building
blocks and can complement theoretical considerations on how to build
such a model. We also emphasize that explicitly infrafred unsafe
-- and in particular infrared dangerous --
observables should complement the tuning of hadronization models and
can be constructed in a way that they are insensitive to hard physics,
and maximally sensitive to hadronization within different ranges of
relevant scales. Our approach will be unified with a similar view on
colour reconnection, based on exploratory steps in \cite{SGE_CR}, and as a
unique consequence of the evolution equations proposed in \cite{Platzer:2022jny}.

Our present analysis is in line with some of the motivations presented
in \cite{Hoang:2024zwl}, and able to generalize this approach towards a more
universal model. We emphasize that the construction of the dynamic
model can benefit from the class of observables we have developed
here, and that we should ultimately seek to construct a new model
which would include gluons and quarks as cluster constituents on the
same footing. In fact, infrared safety in a more general sense as
presented in \cite{Platzer:2022jny}, does imply that this needs to be
the case. In particular, a closer link to re-organizing the low-scale
perturbative Feynman graphs is possible within our paradigm, and has
motivated us to use a 2PI picture ({\it i.e.} one in which two-body
interactions of constituents define propagators of internal systems,
and which will be removed when truncating Green's functions) to
identify the cluster fission matrix elements. Our analysis needs to be
completed by including cluster propagators, and then ultimately allows
us to leverage input from Bethe-Salpeter amplitudes to construct a
model in which more building blocks can be obtained from first
principles.

\section*{Acknowledgments}

We would like to thank Gernot Eichmann, Massimilliano Procura, Daniel
Samitz and Wouter Waalewijn for fruitful discussions. \\ This work has
been supported in part by the BMBF under grant agreements BMBF
05H21VKCCA and 05H24VKB. S.K. acknowledges the support through a KHYS
travel grant and the kind hospitality of the Institute of Physics of
the NAWI Graz.

\bibliography{bib}

\clearpage

\end{document}